\documentclass[10pt]{article}
\usepackage{geometry}
\usepackage{blindtext}
\usepackage{authblk}
\usepackage{graphicx}
\usepackage{dcolumn}
\usepackage{bm}
\usepackage{amsmath}
\usepackage{amssymb}
\usepackage{mathrsfs}
\usepackage{soul}
\usepackage{xcolor}
\usepackage{lineno}

\title{\textbf{Direct stress imaging from shear wave propagation}}

\author[1]{Yiwei Duan}
\author[2,3]{Michel Destrade}
\author[1,4,5,6,*]{Wenchang Tan}
\author[1,*]{Guo-Yang Li}
\affil[1]{School of Mechanics and Engineering Science, Peking University, Beijing 100871, PR China}
\affil[2]{School of Mathematical and Statistical Sciences, University of Galway, University Road, Galway, H91 TK33, Ireland}
\affil[3]{Key Laboratory of Soft Machines and Smart Devices of Zhejiang Province and Department of Engineering Mechanics, Zhejiang University, Hangzhou 310027, People's Republic of China}
\affil[4]{PKU-HKUST Shenzhen-Hong Kong Institution, Shenzhen 518036, China}
\affil[5]{Shenzhen Graduate School, Peking University, Shenzhen 518055, China}
\affil[6]{Shenzhen Bay Laboratory, Shenzhen 518132, China}
\affil[*]{To whom correspondence should be addressed: G.-Y. Li (lgy@pku.edu.cn), W. Tan (tanwch@pku.edu.cn).}
\date{}

\begin{document}

\maketitle
\begin{abstract}
Quantitative imaging of stress fields in heterogeneous solids remains challenging because stress is not directly measurable and is typically inferred from deformation using constitutive models. Here we present Acoustoelastic Imaging (AEI), a non-destructive framework for reconstructing stress fields from shear wave propagation. AEI exploits the acoustoelastic effect, whereby pre-existing stress modifies local wave dynamics, and formulates stress recovery as an inverse problem of the governing wave equations. Using full shear waveform inversion with physics-informed learning, AEI reconstructs wave-equation coefficients from full-field wave measurements, enabling estimation of stress magnitude and principal directions without explicit constitutive model specification or material-parameter calibration. We demonstrate sub-wavelength spatial resolution (\({< 0.28 \lambda}\)) and accurate reconstruction of nonuniform stress fields in heterogeneous materials through numerical simulations and ultrasound shear wave elastography experiments. These results establish a general framework for high-resolution stress imaging and provide a route toward non-invasive mapping of internal mechanical states in complex materials and biological tissues.
\end{abstract}

\clearpage

\section*{Introduction}
Mechanical stress plays a fundamental role in the behavior and function of soft materials, both biological and artificial. 
In living tissues, stress regulates essential cellular processes such as migration~\cite{RN1679,RN1514}, division~\cite{RN669,RN1612}, and mechanosensing~\cite{RN1564,RN708,RN1627}, while at the tissue level, constrained growth and differential forces drive shape changes and mechanical instabilities~\cite{RN1678,RN1546,RN854}. 
Muscle-generated forces underpin nearly all bodily movements~\cite{RN1033,RN1605}, and even at rest, tissues remain mechanically stressed in ways critical to physiological function~\cite{RN414,RN1225}. 
Similarly, in artificial soft materials, internal and applied stresses strongly influence the performance of soft machines, wearable devices, and implantable bioelectronics~\cite{RN979,RN1681,RN1682}. 
Motivated by these observations, developing constitutive models that incorporate initial stress as an additional state variable has become an active area of research, reflecting the growing recognition that pre-existing stress fundamentally alters the mechanical response of soft materials~\cite{RN741,RN1677,RN743}. 
Accurate knowledge of stress distributions is therefore essential for understanding biological processes, assessing pathology, and optimizing the mechanical performance and reliability of soft materials, including fatigue and failure.

Despite its importance, measuring stress in soft materials \emph{in situ} remains a major challenge~\cite{RN1536,RN1027,RN1689}.
Unlike displacement, which can often be observed directly, stress is a tensorial mechanical state variable that is not directly accessible experimentally.
Existing approaches typically infer stress from measured deformations through constitutive models and assumptions about material properties and reference configurations.
Examples include destructive techniques such as hole drilling, as well as non-destructive methods based on X-ray diffraction, neutron diffraction, and ultrasound~\cite{RN1639,RN1643,RN1646}. 
Although successful in specific settings, these approaches generally require material calibration and often become challenging to apply in soft, heterogeneous, or biologically relevant materials. 
Consequently, high-resolution mapping of internal stress fields remains a longstanding challenge in mechanics.

One promising route towards stress measurement exploits the acoustoelastic effect, whereby pre-existing stress alters acoustic wave propagation.~\cite{RN145,RN146}.
Because stress modifies wave speeds and wave trajectories, information about the underlying stress state is encoded in the measured wave field.
Acoustoelasticity has been widely used for nondestructive stress evaluation in engineering materials and has recently attracted growing interest in soft materials and biological tissues. 
However, existing approaches largely rely on local wave-speed measurements, calibrated higher-order elastic constants, or simplified assumptions regarding geometry and material homogeneity. 
As a result, stress imaging with high spatial resolution remains largely inaccessible~\cite{RN51,RN1498}.

Here we introduce Acoustoelastic Imaging (AEI), a framework for reconstructing stress fields directly from shear wave propagation.
The central idea is that stress is encoded in the coefficients of the acoustoelastic wave equations and can therefore be inferred from measured wave fields through inversion. 
By combining full shear waveform inversion with physics-informed learning, AEI reconstructs these coefficients from full-field shear wave measurements, enabling estimation of both stress magnitude and principal directions without explicit constitutive-model specification or material-parameter calibration. 
Because the method leverages the full wave field, including near-field components, it achieves sub-wavelength spatial resolution and remains applicable in elastically heterogeneous materials.

We validate AEI using numerical simulations and experimental measurements acquired by ultrasound shear wave elastography. 
The results demonstrate accurate reconstruction of nonuniform and residual stress fields and show that stress can be separated from material stiffness using the same wave measurements. 
Together, these findings establish a general framework for high-resolution stress imaging and provide a route towards non-invasive characterization of internal mechanical states in complex materials and biological tissues.

\section*{Results}
\subsection*{Stress encoding in shear-wave propagation}

\begin{figure*}[b]
\centering
\includegraphics[width=1\textwidth]{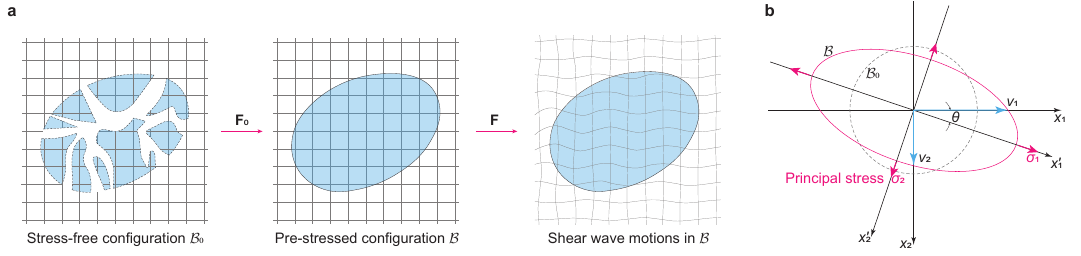}
\caption{Principle of AEI. 
(a) Schematic of the stress-free configuration (\(\mathcal{B}_0\)) and the pre-stressed configuration (\(\mathcal{B}\)), in which shear waves propagate as small-amplitude incremental motions superimposed on the finite stress state.  
(b) Stress-induced anisotropy of shear-wave propagation in \(\mathcal{B}\) under a plane-stress state \((\sigma_1, \sigma_2)\). The principal stress directions determine the anisotropic wave-speed distribution, establishing the basis for recovering stress from wave measurements.}
\label{fig:1}
\end{figure*}

To reconstruct stress from wave measurements, it is first necessary to understand how stress influences shear-wave propagation. The physical basis of AEI is the acoustoelastic effect, whereby pre-existing stress modifies the incremental elastic response of a material and thereby alters the propagation of small-amplitude waves.
Consider a Cauchy stress \(\boldsymbol{\sigma}\) that deforms an elastic body from a stress-free reference configuration \(\mathcal{B}_0\) to the current configuration \(\mathcal{B}\), in which shear waves propagate (Fig.~\ref{fig:1}a). The infinitesimal (incremental) displacement field \(\boldsymbol{u}(\boldsymbol{x}, t)\) in \(\mathcal{B}\) satisfies the incremental equation of motion~\cite{RN105}:
\begin{equation}\label{eq:1}
    \frac{\partial}{\partial x_p}\!\left( \mathcal{A}^0_{piqj}\,\frac{\partial u_j}{\partial x_q} \right)
    - \frac{\partial \hat{\bar{p}}}{\partial x_i}
    + \frac{\partial \bar{p}}{\partial x_j}\frac{\partial u_j}{\partial x_i}
    = \rho\,\frac{\partial^2 u_i}{\partial t^2},
\end{equation}
where \(\rho\) is the mass density in the current configuration, and Einstein summation is adopted throughout. The Eulerian elasticity tensor \(\boldsymbol{\mathcal{A}}^0\) is defined as
\(
\mathcal{A}^0_{piqj} = F_{p\alpha} F_{q\beta}\,\frac{\partial^2 W}{\partial F_{i\alpha}\,\partial F_{j\beta}},
\)
where \(W(\boldsymbol{F})\) is the strain-energy function and \(F_{i\alpha} = \partial x_i / \partial X_\alpha\) is the deformation gradient mapping reference coordinates \(X_\alpha\) to current coordinates \(x_i\) (\(i, \alpha \in \{1,2,3\}\)). For incompressible materials, the Cauchy stress is \(\boldsymbol{\sigma} = \boldsymbol{F}\,\frac{\partial W}{\partial \boldsymbol{F}} - \bar{p}\,\boldsymbol{I}\), where \(\bar{p}\) is the Lagrange multiplier enforcing incompressibility, and \(\hat{\bar{p}}\) in Eq.~\eqref{eq:1} denotes its increment induced by wave motion. Incompressibility further requires \(\partial u_i / \partial x_i = 0\).

A key observation is that the governing wave equation contains elasticity coefficients that are directly linked to stress. 
Specifically, certain components of the Eulerian elasticity tensor satisfy~\cite{RN741}:
\[
\mathcal{A}^0_{ijij} - \mathcal{A}^0_{jiji} = \sigma_{ii} - \sigma_{jj} \quad (\text{no summation}).
\]
Because these components appear explicitly in Eq.~\eqref{eq:1}, the stress difference \((\sigma_{ii} - \sigma_{jj})\) can, in principle, be inferred directly from wave propagation without prior knowledge of the constitutive law. However, this stress difference is coordinate-dependent. Transforming to the principal coordinate system \((x'_1, x'_2, x'_3)\), where the Cauchy stress is diagonal with principal stresses \(\sigma_1, \sigma_2, \sigma_3\), yields
\begin{equation}\label{eq:2}
    \mathcal{A}^0_{i'j'i'j'} - \mathcal{A}^0_{j'i'j'i'} = \sigma_i - \sigma_j .
\end{equation}
Equation~\eqref{eq:2} establishes a direct connection between measurable wave dynamics and the underlying stress state, providing the foundation for stress reconstruction from wave propagation.

To illustrate the implications of Eq.~\eqref{eq:2}, we consider a 2D homogeneous stress state \((\sigma_1, \sigma_2)\) whose principal directions are inclined by an angle \(\theta\) relative to the laboratory coordinate system (Fig.~\ref{fig:1}b). A plane shear wave propagating in the principal coordinate system \((x'_1, x'_2)\) can be expressed as
\(
\boldsymbol{u} = \boldsymbol{u}_0 \exp\!\left[i k \left(x'_1 \cos \vartheta + x'_2 \sin \vartheta - v t\right)\right],
\)
where \(\boldsymbol{u}_0\) is the displacement amplitude, \(\vartheta\) is the propagation angle relative to \(x'_1\), \(v\) is the wave speed, \(k\) is the wavenumber, and \(t\) is time. Substituting this into Eq.~\eqref{eq:1} yields (see \emph{SI Appendix}, Note1)
\begin{equation}\label{eq:3}
\rho v^2
= \alpha \cos^4 \vartheta
+ 2 \beta \cos^2 \vartheta \sin^2 \vartheta
+ \gamma \sin^4 \vartheta ,
\end{equation}
where \(\alpha = \mathcal{A}^0_{1'2'1'2'}\), \(2\beta = \mathcal{A}^0_{1'1'1'1'} + \mathcal{A}^0_{2'2'2'2'} - 2\mathcal{A}^0_{1'1'2'2'} - 2\mathcal{A}^0_{2'1'1'2'}\), and \(\gamma = \mathcal{A}^0_{2'1'2'1'}\).

Equation~\eqref{eq:3} predicts an elliptical shear-wavefront whose principal axes align with the principal stress directions (see \emph{SI Appendix}, Fig.~S1), implying that the orientation angle \(\theta\) can be determined from the wavefront geometry (Fig.~\ref{fig:1}b). Considering two shear waves along the laboratory axes \(x_1\) and \(x_2\) with speeds \(v_1\) and \(v_2\), respectively:
\(
\rho (v_1^2 - v_2^2) = (\alpha - \gamma)\cos 2\theta.
\)
Since \(\alpha - \gamma = \mathcal{A}^0_{1'2'1'2'} - \mathcal{A}^0_{2'1'2'1'} = \sigma_1 - \sigma_2\), the principal stress difference can be obtained from measurable wave speeds:
\begin{equation}\label{eq:4}
\sigma_1 - \sigma_2
= \rho\,\frac{v_1^2 - v_2^2}{\cos 2\theta}.
\end{equation}
Equation~\eqref{eq:4} shows that the principal stress difference is encoded in measurable characteristics of shear-wave propagation, indicating that stress information can, in principle, be recovered directly from wave data.

Real materials generally exhibit spatially varying stress fields, for which local wave-speed measurements become insufficient. Instead, AEI reconstructs the local coefficients of the governing wave equations \(\mathcal{A}^0_{i'j'i'j'}\) from full-field shear-wave measurements and subsequently recovers the corresponding stress distribution.
Assuming small stress gradients, \(\partial \mathcal{A}^0_{p'i'q'j'} / \partial x_q \approx 0\) and \(\partial \bar{p} / \partial x_j \approx 0\), the wave equation simplifies in the principal coordinate system:
\begin{equation}\label{eq:5}
\mathcal{A}^0_{p'i'q'j'}
\frac{\partial^2 u_{j'}}{\partial x_{p'} \partial x_{q'}}
- \frac{\partial \hat{\bar{p}}}{\partial x_{i'}}
= \rho\,\frac{\partial^2 u_{i'}}{\partial t^2}.
\end{equation}
This approximation slightly reduces spatial resolution but significantly lowers computational cost. Once \(\mathcal{A}^0_{p'i'q'j'}\) is inferred, the corresponding stress field can be reconstructed using Eq.~\eqref{eq:2}.

\subsection*{Reconstruction of stress from full-wave inversion}
\begin{figure*}[htb!]
\centering
\includegraphics[width=1\textwidth]{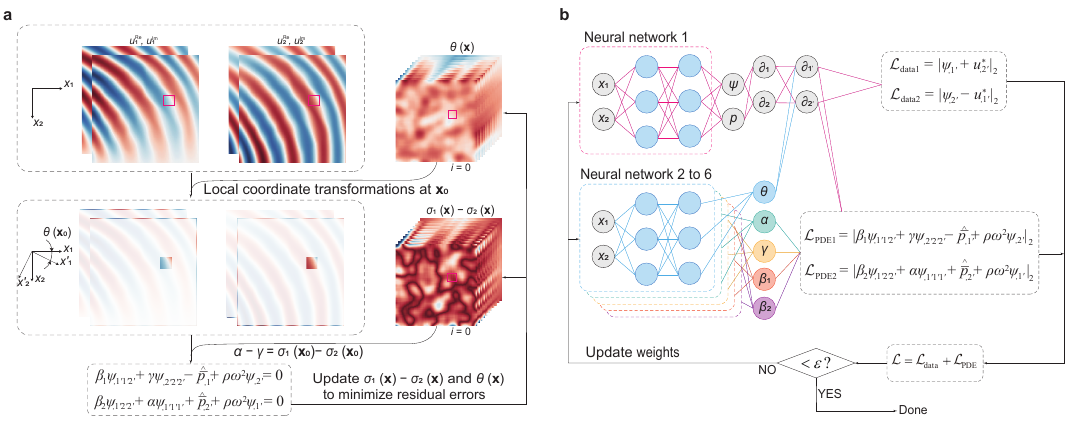}
\caption{FSWI enabled by physics-informed deep learning for AEI.
(a) Concept of FSWI for AEI. The measured displacement field is locally transformed into the principal coordinate system to compute residuals of the wave equation.  
(b) Schematic of the physics-informed neural network (PINN) architecture used for FSWI. Six independent fully connected neural networks are employed to infer displacement fields, principal directions, and Eulerian elasticity coefficients.}
\label{fig:2}
\end{figure*}

The preceding analysis shows that stress information is encoded in the coefficients of the acoustoelastic wave equations. 
Recovering stress from measured wave fields therefore becomes an inverse problem: determining the spatially varying wave-equation coefficients that best explain the observed wave propagation.
To solve this problem, we employ full shear waveform inversion (FSWI), which uses the complete measured wave field rather than local wave-speed estimates. 
Figure~\ref{fig:2}a illustrates the overall reconstruction strategy.
For simplification, we consider 2D harmonic wave motion, where \(\rho \partial^2 () / \partial t^2 = -\rho \omega^2 ()\) and \(\omega\) denotes the angular frequency. 
The displacement components in the laboratory coordinate system are \(u^*_1\) and \(u^*_2\). 
This full shear wave field can be measured using imaging techniques such as ultrasound ~\cite{RN1547}, optical coherence tomography~\cite{RN39}, or magnetic resonance imaging ~\cite{RN764}. 

FSWI seeks the stress field \(\sigma_1(\boldsymbol{x}) - \sigma_2(\boldsymbol{x})\) and principal direction \(\theta(\boldsymbol{x})\) that produce wave fields consistent with both the measurements \(u^*_1\) and \(u^*_2\) and the governing wave equations. At each location \(\boldsymbol{x}_0\), the measured wave field is transformed into the local principal coordinate system, where the stress-dependent coefficients enter the wave equation through Eq.~\eqref{eq:2}. The mismatch between measured and predicted wave behavior is then minimized iteratively until convergence (see Material and Methods).

To implement FSWI efficiently, we develop a physics-informed neural network (PINN) architecture that represents the unknown wave fields and acoustoelastic coefficients as continuous functions of space.
The network consists of six fully connected neural networks (Fig.~\ref{fig:2}b). One network represents the displacement field and \(\hat{\bar{p}}\), while the remaining networks represent the principal directions and wave-equation coefficients. The displacement field is parameterized through a stream function to satisfy incompressibility automatically. Details of the architecture and loss functions are provided in Materials and Methods.

Importantly, the quantities of primary interest—the stress difference \(\bar{\sigma}\) and principal direction \(\theta\)—are uniquely determined by the inversion. In contrast, the individual coefficients \(\alpha\) and \(\gamma\) are identifiable only up to a common additive constant and therefore are not uniquely recoverable (see \emph{SI Appendix}, Note 3). This non-uniqueness does not affect stress reconstruction because the stress depends only on the difference \(\alpha-\gamma\).

\subsection*{Reconstruction of hidden residual stress fields}
\begin{figure*}[t]
\centering
\includegraphics[width=1\textwidth]{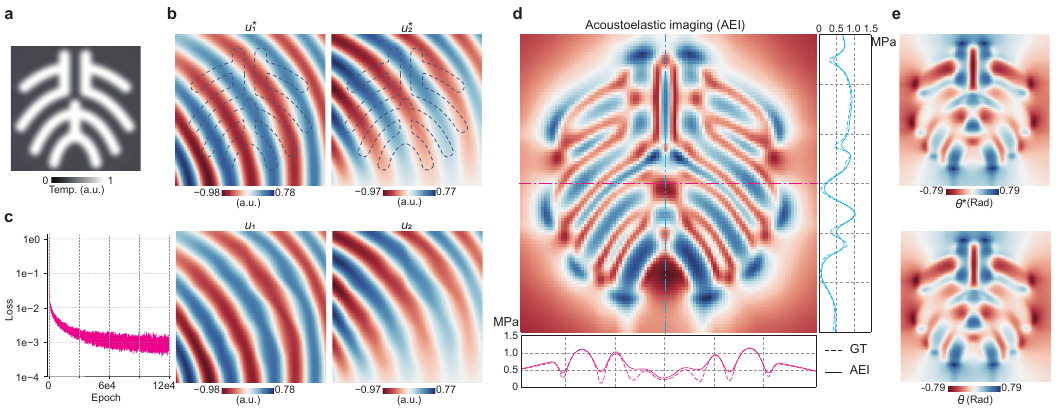}
\caption{Reconstruction of hidden stress fields using AEI in simulated shear-wave data. 
(a) Temperature field used to generate nonuniform thermal stresses in a load-free medium.  
(b) Shear-wave field propagating through the stressed medium, showing wavefront distortions induced by the underlying stress distribution, compared with the reconstructed field obtained from the trained PINN.  
(c) Convergence of the loss function during physics-informed inversion.  
(d) Reconstructed stress profiles along horizontal and vertical centerlines, showing agreement with the ground-truth (GT) distribution.  
(e) Comparison of reconstructed principal stress directions with GT, demonstrating accurate recovery of stress orientation.}
\label{fig:3}
\end{figure*}

Residual stresses provide a stringent test for stress imaging because they are internally self-equilibrated and therefore cannot be inferred from externally applied loads alone. To evaluate AEI under these conditions, we consider a material containing a spatially nonuniform residual stress field generated by constrained thermal expansion (Fig.~\ref{fig:3}a; see Materials and Methods).
Although no external loads are present, the imposed temperature field generates complex internal stresses throughout the material. Harmonic shear waves are introduced at one boundary, and the resulting wave field within a region of interest is used for reconstruction.
The resulting wave field exhibits only subtle deviations from stress-free propagation (Fig.~\ref{fig:3}b), indicating that the stress information is embedded in small distortions of the wavefront. Despite their weak visual appearance, these distortions are accurately captured by the trained PINN, whose reconstructed displacement field closely matches the simulated measurements. The corresponding loss evolution demonstrates stable convergence of the inversion process (Fig.~\ref{fig:3}c).

Applying AEI to the measured wave field enables reconstruction of both the stress difference \(\bar{\sigma}\) and the principal direction \(\theta\) (Figs.~\ref{fig:3}d and e). The reconstructed fields accurately reproduce the spatial patterns and orientations of the ground-truth solution, demonstrating that complex residual stresses can be recovered from wave propagation alone.
Quantitative comparison along the horizontal and vertical centerlines confirms excellent agreement between reconstructed and ground-truth stresses (Fig.~\ref{fig:3}d). Small discrepancies appear primarily in regions with steep stress gradients, where the local stress-gradient approximation (\(\partial \mathcal{A}^0_{p'i'q'j'} / \partial x_q \approx 0\)) employed in the inversion becomes less accurate. Incorporating the full wave equation could further improve reconstruction in such regions, albeit with increased computational cost.

We further compare the reconstructed coefficients \(\alpha\) and \(\gamma\) with their ground-truth values (see \emph{SI Appendix}, Fig.~S2). Consistent with the theoretical analysis, the reconstructed stress and principal directions are unique, whereas the individual coefficients are not. Specifically, \(\alpha\) and \(\gamma\) differ from the ground truth by an approximately uniform offset while preserving their difference, which uniquely determines the stress field.

Because AEI relies on identifying stress-induced perturbations of the wave field, its performance is expected to depend on measurement noise. To quantify this sensitivity, we introduce synthetic noise at different levels and repeat the reconstruction. The accuracy of the recovered stress field exhibits a strong dependence on signal-to-noise ratio (SNR). For the cases considered here, stress reconstruction remains accurate for SNR values exceeding approximately \(30\) dB (see \emph{SI Appendix}, Fig.~S3), indicating that AEI is compatible with the noise levels achievable in modern wave-based imaging systems~\cite{RN39}.

\subsection*{Sub-wavelength resolution in stress imaging}

\begin{figure*}[htb!]
\centering
\includegraphics[width=1\textwidth]{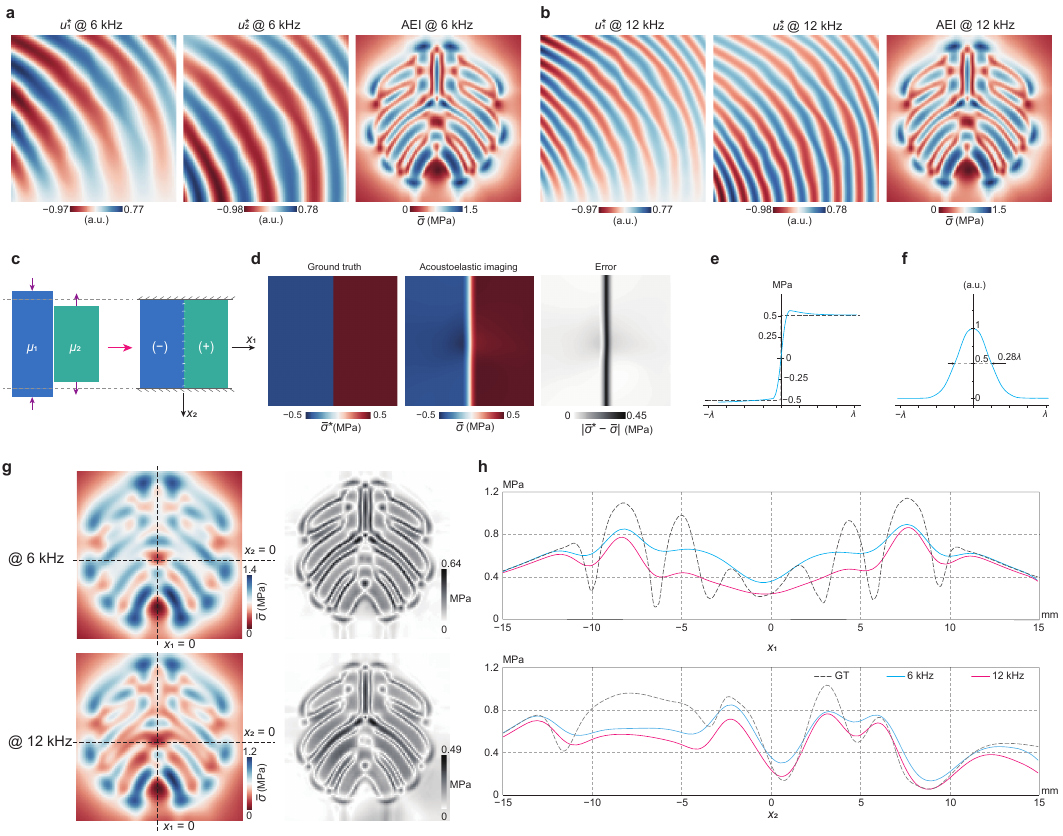}
\caption{Sub-wavelength resolution in stress imaging with AEI. 
(a,b) Stress reconstructions obtained using 6 kHz and 12 kHz shear waves, respectively, showing that the recovered stress field is largely invariant with respect to excitation frequency.  
(c) Schematic of a sharp step-transition stress configuration used to quantify spatial resolution.  
(d) Comparison between the reconstructed stress and the ground truth (GT) for the step-transition case.  
(e) Cross-sectional stress profile across the interface, showing a finite transition zone in the reconstructed stress compared with the idealized discontinuity.  
(f) Derivative of the stress profile used to quantify resolution, with the full width at half maximum (FWHM) yielding a spatial resolution of approximately \(0.28\lambda\).  
(g) Reconstruction using only the vertical displacement component (\(u_2^*\)), illustrating the effect of incomplete wave-field information.  
(h) Comparison between reconstruction with incomplete data and GT, showing reduced accuracy and spatial resolution when wave-field components are missing.}
\label{fig:4}
\end{figure*}

A notable feature of the reconstructed stress maps is their ability to resolve fine spatial details that are substantially smaller than the shear wavelength (Fig.~\ref{fig:3}d). This observation suggests that AEI can achieve sub-wavelength resolution. To investigate this capability, we first examine the influence of shear-wave frequency on stress reconstruction. Doubling the excitation frequency from 6 kHz to 12 kHz reduces the wavelength by a factor of two, yet the reconstructed stress patterns remain nearly unchanged (Figs.~\ref{fig:4}a and b). In particular, stress features that are sub-wavelength at 6 kHz are still accurately recovered. This weak dependence on wavelength contrasts with conventional far-field imaging approaches, whose spatial resolution is fundamentally constrained by diffraction.

To quantify the achievable resolution, we consider a stress field containing a sharp step transition. Two media subjected to uniaxial compression and tension, respectively, are joined together, and shear-wave propagation across the interface is simulated (Fig.~\ref{fig:4}c). Applying AEI yields the reconstructed stress field shown in Fig.~\ref{fig:4}d, while the inferred principal directions are provided in \emph{SI Appendix}, Fig.~S3. Although the true stress distribution contains a discontinuity, the reconstructed stress exhibits a finite transition zone (Fig.~\ref{fig:4}e). We characterize the spatial resolution by computing the full width at half maximum (FWHM) of the derivative of the reconstructed stress profile. The resulting resolution is approximately \(0.28\lambda\) (Fig.~\ref{fig:4}f), demonstrating sub-wavelength stress imaging well beyond the diffraction limit associated with conventional far-field wave imaging.

The origin of this resolution enhancement lies in the use of the full measured shear-wave field. Unlike approaches based on local wave-speed estimation, AEI reconstructs stress by fitting the complete wave field to the governing acoustoelastic equations. Consequently, both propagating and near-field wave components contribute to the inversion. Similar super-resolution behavior has been reported in full-waveform inversion for shear wave elastography (SWE)~\cite{RN497}, where sub-wavelength elastic heterogeneities can be resolved by exploiting near-field wave distortions. In AEI, the near-field wave components encode subtle perturbations induced by stress, enabling reconstruction of stress variations on length scales substantially smaller than the wavelength.

Because AEI relies on information contained in the full vector wave field, incomplete wave measurements are expected to degrade reconstruction performance. To evaluate this effect, we repeat the inversion using only the vertical displacement component \( u_2^*\), which is often the only component available in practical imaging systems. The reconstructed stress maps are shown in Figs.~\ref{fig:4}g and d. As expected, the loss of wave-field information reduces both reconstruction accuracy and spatial resolution. Under these conditions, increasing the excitation frequency leads to improved spatial resolution, in contrast to the frequency-independent performance observed when the full wave field is available (Figs.~\ref{fig:4}a and b). These results suggest that higher-frequency shear waves can partially compensate for incomplete measurements and improve the practical performance of AEI.

\subsection*{Separating stress from material stiffness}

In practical materials and biological tissues, wave propagation is influenced by both stress and elastic heterogeneity. Distinguishing these effects is challenging because each modifies the measured wave field. A critical question is therefore whether stress can be reconstructed independently of spatial variations in material stiffness.
To address this question, we consider a heterogeneous material containing a stiff inclusion embedded within a softer matrix together with a nonuniform stress field (see Materials and Methods). In this configuration, wave propagation is affected simultaneously by variations in stiffness and stress, providing a stringent test for AEI.

\begin{figure*}[t]
\centering
\includegraphics[width=1\textwidth]{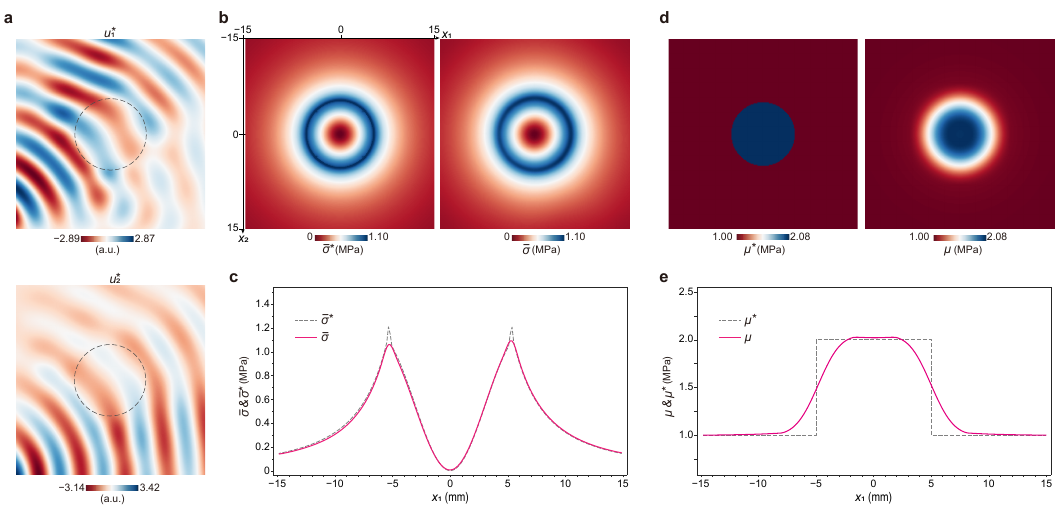}
\caption{Separation of stress and stiffness in heterogeneous materials using AEI. 
(a) Shear-wave field in a heterogeneous soft solid consisting of a compliant matrix and a stiffer circular inclusion. A nonuniform stress field is generated by constrained growth of the inclusion.  
(b) Stress field reconstructed by AEI (\(\bar{\sigma}\)) compared with the ground truth (\(\bar{\sigma}^*\)).  
(c) Line profiles of the reconstructed stress along the \(x_1\)-axis showing quantitative agreement with the ground truth.  
(d) Shear modulus map obtained using shear wave elastography (SWE) under the assumption of a stress-free material, compared with the ground truth (\(\mu^*\)).  
(e) Line profiles of the reconstructed shear modulus along the \(x_1\)-axis.}
\label{fig:5}
\end{figure*}

Applying AEI to the simulated wave field yields (Fig.~\ref{fig:5}a) the reconstructed stress distribution and principal directions shown in Figs.~\ref{fig:5}b and c. Despite the presence of substantial elastic heterogeneity, the reconstructed stress field agrees closely with the ground truth. These results demonstrate that AEI can recover stress information from the wave field even when stiffness varies spatially.

To place these results in context, we analyze the same wave measurements using shear wave elastography (SWE), which interprets wave propagation primarily in terms of material stiffness.
Applying SWE to the wave field under the assumption of a stress-free material yields the shear-modulus map shown in Fig.~\ref{fig:5}d. The stiff inclusion is correctly identified and the reconstructed modulus agrees well with the ground truth. However, the modulus map exhibits a gradual transition near the inclusion boundary. This apparent smoothing arises because stress-induced changes in wave propagation are partially interpreted as variations in stiffness.

These results highlight a fundamental distinction between AEI and conventional elastography. SWE aims to characterize material properties, whereas AEI reconstructs the mechanical state of the material. Although both approaches rely on the same wave measurements, they recover different physical quantities. Consequently, AEI and SWE provide complementary information: one characterizes stiffness, while the other characterizes stress.

\subsection*{Experimental validation with ultrasound shear wave elastography}

\begin{figure*}[t]
\centering
\includegraphics[width=1\textwidth]{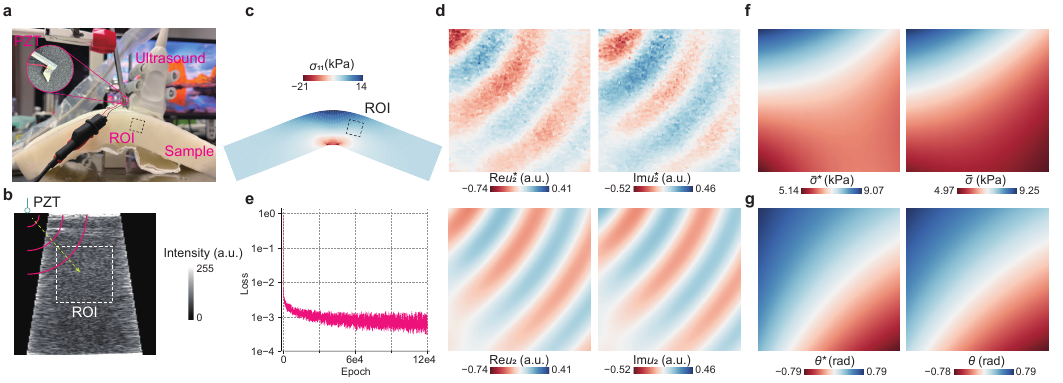}
\caption{Experimental implementation of AEI using ultrasound shear wave elastography (SWE). 
(a) Photograph of the experimental setup.  
(b) Grayscale ultrasound image of the PVA hydrogel sample, with the dashed square indicating the region of interest (ROI) used for AEI reconstruction.  
(c) Reference stress field obtained from finite element analysis under the same boundary and loading conditions.  
(d) Measured shear-wave field (real and imaginary components) within the ROI and comparison with the reconstructed field obtained from physics-informed inversion.  
(e) Convergence of the inversion process, shown by the evolution of the loss function.  
(f,g) Reconstructed stress field and principal stress directions obtained from AEI, compared with finite-element reference results, demonstrating quantitative agreement under experimental conditions.}
\label{fig:6}
\end{figure*}

To evaluate the practical feasibility of AEI, we implemented the method using ultrasound shear wave elastography (SWE), a clinically established platform for measuring shear-wave propagation in soft tissues. Figure~\ref{fig:6}a shows the experimental setup, where shear waves are generated using a PZT actuator (500 Hz) and recorded using ultrafast ultrasound imaging (Fig.~\ref{fig:6}b), providing the full-field wave data required for AEI reconstruction.

To provide a reference for validation, we constructed a finite element model with identical boundary and loading conditions, using a Young’s modulus estimated from shear-wave speed measured in the stress-free sample (see \emph{SI Appendix}, Fig.~S5). The resulting simulation yields a nonuniform stress field that serves as ground truth (Fig.~\ref{fig:6}c).

Although the imaging field of view is limited, AEI is applied to the measured wave field within a region of interest. The acquired shear waves exhibit only weak apparent distortions, with a modest increase in average wave velocity from \(4.3\)~m/s (stress-free state) to \(4.5\)~m/s, and no visually obvious signatures of stress heterogeneity (Fig.~\ref{fig:6}d). No spatial filtering is applied prior to inversion, as the physics-informed formulation intrinsically regularizes noise.

Applying AEI enables reconstruction of both the stress field and principal directions from the experimental data. The inferred stress distribution agrees well with finite element predictions, while the corresponding loss function shows stable convergence with a reduction of three orders of magnitude (Figs.~\ref{fig:6}e and f). The largest deviations occur near the edge of the imaging domain, where signal-to-noise ratio is reduced due to geometric spreading of the shear waves.

These results demonstrate that AEI can be implemented using standard ultrasound SWE platforms and is capable of extracting internal stress fields from experimentally measured wave data under realistic imaging conditions.

\section*{Discussion}

Quantitative characterization of internal stress fields in solids remains a central challenge in mechanics. The difficulty is fundamental: stress is not directly measurable and is typically inferred indirectly from displacement fields through constitutive assumptions. In heterogeneous or nonlinear materials, this indirectness limits both accuracy and generality, and constrains our ability to access internal mechanical states noninvasively.

Here we introduce Acoustoelastic Imaging (AEI), a framework for reconstructing stress fields directly from shear-wave propagation. The central idea is that pre-existing stress modifies local wave dynamics through the acoustoelastic effect, embedding information about the internal stress state into the coefficients of the governing wave equation. By solving the resulting inverse problem using physics-constrained waveform inversion, AEI enables reconstruction of both stress magnitude and principal directions without requiring an explicit constitutive model or prior material calibration.

A key implication of this formulation is that stress can be treated as a measurable mechanical state variable encoded in wave propagation. This contrasts with conventional shear wave elastography, which interprets wave fields primarily in terms of material stiffness. In AEI, the same wave data is reinterpreted through a different physical lens, revealing stress as an independent observable rather than a secondary inference. Our results further demonstrate that stress and stiffness contribute differently to wave propagation and can be separated in heterogeneous materials, establishing AEI and elastography as complementary modalities probing distinct mechanical attributes.

Beyond performance, AEI suggests a broader conceptual shift in how mechanical information is extracted from wave fields. While displacement is a kinematic field, stress is a tensorial state variable governing internal equilibrium. The ability to reconstruct stress directly therefore represents a transition from kinematic imaging to tensorial state-variable imaging in continuum mechanics. This perspective naturally generalizes beyond stress alone, suggesting a broader class of inverse problems in which internal tensor fields are inferred directly from wave propagation.

A natural extension of the present framework is its generalization to compressible media. While the current formulation is developed and validated in incompressible soft solids for experimental clarity, the underlying principle is more general: wave propagation encodes the acoustoelastic response of the underlying mechanical state through the governing wave tensor. In compressible materials, volumetric deformation introduces additional coupling between stress components and wave kinematics, thereby enriching the structure of the wave-equation coefficients. Accounting for these effects would extend AEI to a broader class of solids, including biological tissues, porous media, and engineered composites, where compressibility plays a non-negligible role in the internal mechanical state.

Several limitations remain. The present formulation assumes relatively smooth stress variations and simplified wave-field representations, which may affect accuracy in regions with strong heterogeneity. Extending the framework to fully general wave physics, including stronger nonlinearities and multimodal wave coupling, will be important for further development. In addition, experimental validation is currently limited to controlled ultrasound shear-wave measurements, and further studies are required to evaluate performance in complex biological and in vivo environments.

Despite these limitations, AEI establishes a physically grounded framework for reconstructing internal stress fields from wave propagation. More broadly, it provides a foundation for non-invasive mechanical state imaging in soft matter, functional materials, and biological tissues, and opens a route toward tensorial reconstruction of internal states in continuum systems.

\section*{Methods}
\subsection*{Acoustoelastic wave equation for AEI}
We derive the wave equation for AEI based on the simplified incremental equation of motion (Eq.~\eqref{eq:5} in the main text).
We consider two-dimensional harmonic plane wave propagation in the \(x_1\)–\(x_2\) plane. 
To automatically satisfy the incompressibility constraint, a scalar stream function \(\psi(x_1, x_2)\) is introduced such that
\begin{equation}\label{eq:6}
{u_{1'}} = \frac{{\partial \psi }}{{\partial {x_{2'}}}}\mathrm{e}^{\mathrm{i}\omega t}, \quad 
{u_{2'}}= - \frac{{\partial \psi }}{{\partial {x_{1'}}}}\mathrm{e}^{\mathrm{i}\omega t},
\end{equation}
Where \(\mathrm{i} = \sqrt{-1}\), \(\omega\) is the angular frequency, and \(t\) denotes the time.
Substituting Eq.~\eqref{eq:6} into Eq.~\eqref{eq:5} yields the coupled system
\begin{subequations}\label{eq:7}
\begin{align}
&{\beta _1}{\psi _{,1'1'2'}} + \gamma {\psi _{,2'2'2'}} - {\hat{\bar{p}}_{,1'}} + \rho {\omega ^2}{\psi _{,2'}} = 0, \label{eq:7a} \\
&{\beta _2}{\psi _{,1'2'2'}} + \alpha {\psi _{,1'1'1'}} + {\hat{\bar{p}}_{,2'}} + \rho {\omega ^2}{\psi _{,1'}} = 0. \label{eq:7b}
\end{align}
\end{subequations}
The coefficients are defined as
\begin{subequations}\label{eq:8}
\begin{align}
&\alpha  = {\mathcal{A}_{1'2'1'2'}^0}, \\
&\gamma  = {\mathcal{A}_{2'1'2'1'}^0}, \\
&{\beta _1} = {\mathcal{A}_{1'1'1'1'}^0} - {\mathcal{A}_{2'1'1'2'}^0} - {\mathcal{A}_{1'1'2'2'}^0},\\
&{\beta _2} = {\mathcal{A}_{2'2'2'2'}^0} - {\mathcal{A}_{2'1'1'2'}^0} - {\mathcal{A}_{1'1'2'2'}^0}.
\end{align}    
\end{subequations}
The first two coefficients in Eq.~\eqref{eq:8} are related to the deviatoric stress through
\begin{equation}\label{eq:9}
{\sigma _1} - {\sigma _2} = \alpha  - \gamma.
\end{equation}

Equations~\eqref{eq:7a} and~\eqref{eq:7b} are written in the principal coordinate system.
However, in AEI the full field shear waves are expressed in the global lab coordinate system.
We therefore introduce a coordinate rotation from the lab coordinate system to the local principal coordinate system, in which the wave equations are formulated.
The coordinate rotation matrix \(\boldsymbol{R}\) is given by
\begin{equation}\label{eq:10}
\boldsymbol{R} =
\begin{bmatrix}
{\cos \theta } & {\sin \theta } \\
{ - \sin \theta } & {\cos \theta }
\end{bmatrix},
\quad 
\theta \in \left[ 0, \frac{\pi}{2} \right),
\end{equation}
where \(\theta\) denotes the rotation angle from the lab coordinate system to the principal coordinate system.
Accordingly, the transformed coordinates and displacement components are given by
\begin{subequations}
\begin{align}
x_{1'} &= x_1\cos \theta + x_2\sin \theta, \\
x_{2'} &= -x_1\sin \theta + x_2\cos \theta.    
\end{align}    
\end{subequations}
and
\begin{subequations}
\begin{align}
{u_{1'}} = {u_1}\cos \theta + {u_2}\sin \theta, \\
{u_{2'}} = -{u_1}\sin \theta + {u_2}\cos \theta.    
\end{align}    
\end{subequations}

With the aid of this coordinate transformation, the derivatives of \(\psi\) and \(\hat{\bar{p}}\) in the global coordinate system can be related to those in the principal coordinate system, as required in Eq.~\eqref{eq:7}.
The explicit expressions for these derivatives are provided in \emph{SI Appendix}, Note~2.

\subsection*{Neural networks for FSWI}
For incompressible materials, six neural networks are employed in this study. The spatial coordinates \(x_1\) and \(x_2\) are used as inputs. In the primary network, DNN 1, the outputs are the stream function \(\psi\) and the Lagrange multiplier \(\hat{\bar{p}}\). In the remaining networks, the outputs correspond to the elastic constants and the principal directions. After training, the outputs of the neural networks are constrained to match the data and satisfy the physical constraints, namely the wave equation. As a result, DNN 1 accurately predicts \(\psi\) and \(\hat{\bar{p}}\), while the elastic constants and principal directions are obtained from DNNs 2–6.

The network architecture adopted in this work consists of fully connected feedforward neural networks. DNN 1 contains 10 hidden layers with 40 neurons in each layer. DNNs 2–6 each contain 6 hidden layers, with 40 neurons per layer. The activation function used in this study is ReLU. The weights and biases are optimized through training using the backpropagation algorithm, which minimizes the loss function \({\mathcal{L}}\).

In AEI, the loss function \(\mathcal{L}\) consists of two parts: a data-driven term and a physics-informed term. The stream function \(\psi\) is the output of DNN 1, from which the displacement field can be obtained via automatic differentiation. The data-driven loss is defined as \({\mathcal{L}}_{\mathrm{data}} = {\mathcal{L}}_{\mathrm{data1}} + {\mathcal{L}}_{\mathrm{data2}}\), where
\begin{subequations}
\begin{align}\label{eq:13}
&{\mathcal{L}}_{\mathrm{data1}} = \left| \psi_{,1'} + u_{2'}^* \right|_2,\\
&{\mathcal{L}}_{\mathrm{data2}} = \left| \psi_{,2'} - u_{1'}^* \right|_2,
\end{align}    
\end{subequations}
and \(u_{1'}^*\) and \(u_{2'}^*\) denote the measured displacement data. \(|\;\cdot\;|_2\) represents the L2 norm.

The physics-informed part is \({\mathcal{L}}_{\mathrm{PDE}} = {\mathcal{L}}_{\mathrm{PDE1}} + {\mathcal{L}}_{\mathrm{PDE2}}\), where
\begin{subequations}\label{eq:14}
\begin{align}
{\mathcal{L}}_{\mathrm{PDE1}} &= \left| \beta_1 \psi_{,1'1'2'} + \gamma \psi_{,2'2'2'} - \hat{\bar{p}}_{,1'} + \rho \omega^2 \psi_{,2'} \right|_2,\\
{\mathcal{L}}_{\mathrm{PDE2}} &= \left| \beta_2 \psi_{,1'2'2'} + \alpha \psi_{,1'1'1'} + \hat{\bar{p}}_{,2'} + \rho \omega^2 \psi_{,1'} \right|_2.
\end{align}    
\end{subequations}

To balance the data-driven and physics-informed components of the loss function, we define
\begin{equation}\label{eq:15}
\mathcal{L} = \mathcal{L}_{\mathrm{data}} + \lambda \cdot \mathcal{L}_{\mathrm{PDE}},
\end{equation}
where \(\lambda\) is a hyperparameter to be optimized. During training, better convergence is achieved when \(\mathcal{L}_{\mathrm{PDE}}\) remains approximately one order of magnitude smaller than \(\mathcal{L}_{\mathrm{data}}\), particularly in the presence of noisy data.

The number of epochs required for convergence mainly depends on the network size, batch size, learning rate (default 0.001), and other training parameters. For the networks used in this study, we observe that when the number of epochs exceeds 120,000, the loss function gradually approaches its converged value, although some oscillations are observed near convergence. During training, for incompressible cases, training is terminated at 120,000 epochs; for compressible cases, it is terminated at 160,000 epochs.

To perform SWE-based inversion of the shear modulus in a heterogeneous medium, the AEI neural network is modified accordingly.  
The networks associated with the parameters \(\theta\), \(\gamma\), \(\beta_1\), and \(\beta_2\) are no longer used.  
Instead, the principal direction is fixed by setting \(\theta = 0\).  
The parameters \(\gamma\), \(\beta_1\), and \(\beta_2\) are prescribed as \(\alpha\), \(-\alpha\), and \(-\alpha\), respectively, with \(\alpha = \mu\), where \(\mu\) is the shear modulus.  
With these modifications, the AEI framework reduces to the SWENet formulation~\cite{RN872} previously proposed for SWE.

\subsection*{Simulations of shear waves}
Finite element simulations are employed to generate synthetic data for the validation of AEI. All simulations are performed using Abaqus/Standard (Abaqus 2021, Dassault Systèmes®). A sufficiently large two-dimensional computational domain of \(200 \times 200\;{\mathrm{mm}}^2\) is created. To suppress boundary reflections, infinite elements are applied along all boundaries.

The material behavior is modeled using both incompressible and compressible neo-Hookean constitutive models. In all cases, the material density is set to 1000 \({\mathrm{kg/m^3}}\). The shear modulus of the matrix material is 1 MPa. For cases involving a heterogeneous inclusion, the shear modulus of the stiff inclusion core is set to 2 MPa.

The ROI is defined as a subdomain fully illuminated by the propagating shear waves. In the simulations, the ROI size is chosen as \(30 \times 30\;{\mathrm{mm}}^2\).

For complex non-uniform stress fields, non-uniform prestress is generated by prescribing a mapped temperature field within the ROI. The coefficient of thermal expansion of the material is set to \(2.6 \times 10^{-4}\;{\rm K^{-1}}\). The reference temperature within the ROI is \(0^\circ\)C, and the maximum temperature reaches \(1000^\circ\)C, allowing the induced thermal stress to be of the same order of magnitude as the shear modulus.

The simulations are performed in the frequency domain. To generate shear waves, an acoustic radiation force \(f\) acting in the vertical direction is applied in the steady-state dynamic analysis step, given by
\begin{equation}\label{eq:16}
f = f_0 \exp \left[ - \frac{(x_1 - x_{10})^2}{2 r_1^2} - \frac{(x_2 - x_{20})^2}{2 r_2^2} \right].
\end{equation}

In this expression, \(f_0\) specifies the peak magnitude of the body force, while \(r_1\) and \(r_2\) define its spatial distribution. The coordinates \(x_1\) and \(x_2\) denote the location of the applied force. The acoustic radiation force is applied at the lower-left corner of the ROI to illuminate the entire subdomain.

A uniform mesh with an element size of approximately \(0.25 \times 0.25\;{\rm mm^2}\) is employed within the ROI. The element type used in Abaqus is CPE8RH (plane strain, 8-node biquadratic, hybrid with linear pressure, reduced integration). Mesh convergence is verified by progressively reducing the element size, and the variation in results is found to be less than 1\%. After the simulations are completed, the values of \(u_1\) and \(u_2\) at each mesh node within the ROI are extracted and used as input data for the neural networks.

The heterogeneous materials example in this study is implemented in Abaqus by assigning different Young’s moduli to different regions. Within the ROI, there is a circular inclusion with a radius of 5 mm. The Young’s modulus inside the inclusion is 2.08 MPa, while outside the inclusion it is 1 MPa. The diameter of the inclusion is comparable to the shear wave field.

A Gaussian-distributed heat source with a FWHM of 8.33 mm is applied inside the inclusion. The peak temperature of the heat source is \(100^\circ\)C, and the reference temperature is \(0^\circ\)C. This causes the inclusion to expand while being constrained by the surrounding medium, thereby generating self-equilibrated thermal stress within the ROI.

\subsection*{Ultrasound shear wave elastography}
The ultrasound system used in this study was built based on the Vantage 64 LE system (Verasonics Inc., Kirkland, WA, USA).
The ultrasound transducer (L9-4, Shenzhen Jiarui Electronic Technology Co., Ltd., Shenzhen, China) used in the experiment had a center frequency of 7 MHz, a pitch of 0.3 mm, and 128 elements.
Elastic waves were generated on the surface of the hydrogel using a PZT actuator.
During wave propagation, the transducer acquired in-phase and quadrature (IQ) signals at a frame rate of 10 kHz.
Plane-wave imaging with delay-and-sum beamforming was used to reconstruct each frame.
Five consecutive measurements were performed, and the results were averaged to improve the SNR ratio.

\subsection*{Preparation of PVA hydrogel}
The PVA hydrogel consists of \(10\) wt\% PVA, \(3\) wt\% cellulose, and \(87\) wt\% deionized water.
The PVA powder (Sigma-Aldrich, 341584, Shanghai, China) was first dissolved in water at \(80^\circ\)C.
Subsequently, cellulose powder (Sigma-Aldrich, S3504, Shanghai, China) was added to the solution and thoroughly stirred to obtain a suspension containing cellulose particles.
The cellulose particles act as ultrasonic scatterers to enhance imaging contrast.
The suspension was then poured into a rectangular plastic container (approximately \(30\) cm in length, \(8\) cm in width, and \(6\) cm in height), cooled to room temperature (about \(20^\circ\)C), and subsequently placed in a freezer at \(-20^\circ\)C.
The sample was frozen for \(12\) hours and then thawed at room temperature for another \(12\) hours.
The stiffness of the sample can be tuned by adjusting the number of freezing/thawing (F/W) cycles.
In this study, the PVA hydrogel samples underwent three F/W cycles.
\subsection*{Data Availability}
The data supporting the findings of this study are available within the Article and its Supplementary Information. Source data are provided with this paper. Any other information is available from the corresponding authors upon request.

\subsection*{Code Availability}
The code used in this study is available from the corresponding author upon request.

\bibliographystyle{unsrt}
\bibliography{references}

@article{RN146,
   author = {Li, Guo-Yang and Gower, Artur L. and Destrade, Michel},
   title = {An ultrasonic method to measure stress without calibration: The angled shear wave method},
   journal = {The Journal of the Acoustical Society of America},
   volume = {148},
   number = {6},
   pages = {3963-3970},
   ISSN = {0001-4966},
   DOI = {10.1121/10.0002959},
   url = {https://doi.org/10.1121/10.0002959},
   year = {2020},
   type = {Journal Article}
}

@article{RN51,
   author = {Zhang, Zhaoyi and Li, Guo-Yang and Jiang, Yuxuan and Zheng, Yang and Gower, Artur L. and Destrade, Michel and Cao, Yanping},
   title = {Noninvasive measurement of local stress inside soft materials with programmed shear waves},
   journal = {Science Advances},
   volume = {9},
   number = {10},
   pages = {eadd4082},
   DOI = {doi:10.1126/sciadv.add4082},
   url = {https://www.science.org/doi/abs/10.1126/sciadv.add4082},
   year = {2023},
   type = {Journal Article}
}

@inbook{RN105,
   author = {Ogden, Ray W.},
   title = {Incremental Statics and Dynamics of Pre-Stressed Elastic Materials},
   booktitle = {Waves in Nonlinear Pre-Stressed Materials},
   editor = {Destrade, Michel and Saccomandi, Giuseppe},
   publisher = {Springer Vienna},
   address = {Vienna},
   pages = {1-26},
   ISBN = {978-3-211-73572-5},
   DOI = {10.1007/978-3-211-73572-5_1},
   url = {https://doi.org/10.1007/978-3-211-73572-5_1},
   year = {2007},
   type = {Book Section}
}

@article{RN741,
   author = {Shams, M. and Destrade, M. and Ogden, R. W.},
   title = {Initial stresses in elastic solids: Constitutive laws and acoustoelasticity},
   journal = {Wave Motion},
   volume = {48},
   number = {7},
   pages = {552-567},
   ISSN = {0165-2125},
   DOI = {https://doi.org/10.1016/j.wavemoti.2011.04.004},
   url = {https://www.sciencedirect.com/science/article/pii/S0165212511000436},
   year = {2011},
   type = {Journal Article}
}

@article{RN39,
   author = {Feng, Xu and Li, Guo-Yang and Yun, Seok-Hyun},
   title = {Ultra-wideband optical coherence elastography from acoustic to ultrasonic frequencies},
   journal = {Nature Communications},
   volume = {14},
   number = {1},
   pages = {4949},
   ISSN = {2041-1723},
   DOI = {10.1038/s41467-023-40625-y},
   url = {https://doi.org/10.1038/s41467-023-40625-y},
   year = {2023},
   type = {Journal Article}
}

@article{RN764,
   author = {Sack, Ingolf},
   title = {Magnetic resonance elastography from fundamental soft-tissue mechanics to diagnostic imaging},
   journal = {Nature Reviews Physics},
   volume = {5},
   number = {1},
   pages = {25-42},
   ISSN = {2522-5820},
   DOI = {10.1038/s42254-022-00543-2},
   url = {https://doi.org/10.1038/s42254-022-00543-2},
   year = {2023},
   type = {Journal Article}
}

@article{RN1547,
   author = {Zhang, Zhaoyi and Ma, Shiyu and Yin, Ziying and Qiu, Jing and Hu, Zhongtao and Li, Guo-Yang and Feng, Xi-Qiao and Cao, Yanping},
   title = {Unveiling hidden features of acoustic radiation forces in soft tissues via physics-informed neural network-based full shear wave inversion},
   journal = {Journal of the Mechanics and Physics of Solids},
   volume = {205},
   pages = {106326},
   ISSN = {0022-5096},
   DOI = {https://doi.org/10.1016/j.jmps.2025.106326},
   url = {https://www.sciencedirect.com/science/article/pii/S0022509625003023},
   year = {2025},
   type = {Journal Article}
}

@article{RN497,
   author = {Zemzemi, C. and Zorgani, A. and Daunizeau, L. and Belabhar, S. and Souchon, R. and Catheline, S.},
   title = {Super-resolution limit of shear-wave elastography},
   journal = {Europhysics Letters},
   volume = {129},
   number = {3},
   pages = {34002},
   ISSN = {0295-5075},
   DOI = {10.1209/0295-5075/129/34002},
   url = {https://dx.doi.org/10.1209/0295-5075/129/34002},
   year = {2020},
   type = {Journal Article}
}

@article{RN1646,
   author = {Chen, Shu-guang and Gao, Han-jun and Zhang, Yi-du and Wu, Qiong and Gao, Zi-han and Zhou, Xin},
   title = {Review on residual stresses in metal additive manufacturing: formation mechanisms, parameter dependencies, prediction and control approaches},
   journal = {Journal of Materials Research and Technology},
   volume = {17},
   pages = {2950-2974},
   ISSN = {2238-7854},
   DOI = {https://doi.org/10.1016/j.jmrt.2022.02.054},
   url = {https://www.sciencedirect.com/science/article/pii/S2238785422002241},
   year = {2022},
   type = {Journal Article}
}

@article{RN414,
   author = {Nia, Hadi T. and Munn, Lance L. and Jain, Rakesh K.},
   title = {Physical traits of cancer},
   journal = {Science},
   volume = {370},
   number = {6516},
   pages = {eaaz0868},
   DOI = {doi:10.1126/science.aaz0868},
   url = {https://www.science.org/doi/abs/10.1126/science.aaz0868},
   year = {2020},
   type = {Journal Article}
}

@article{RN1677,
   author = {Gower, Artur L. and Shearer, Tom and Ciarletta, Pasquale and Destrade, Michel},
   title = {The elastic stored energy of initially strained, or stressed, materials: restrictions and third-order expansions},
   journal = {Proceedings of the Royal Society A: Mathematical, Physical and Engineering Sciences},
   volume = {481},
   number = {2307},
   ISSN = {1364-5021},
   DOI = {10.1098/rspa.2024.0272},
   url = {https://doi.org/10.1098/rspa.2024.0272},
   year = {2025},
   type = {Journal Article}
}

@article{RN1643,
   author = {Guo, Jiang and Fu, Haiyang and Pan, Bo and Kang, Renke},
   title = {Recent progress of residual stress measurement methods: A review},
   journal = {Chinese Journal of Aeronautics},
   volume = {34},
   number = {2},
   pages = {54-78},
   ISSN = {1000-9361},
   DOI = {https://doi.org/10.1016/j.cja.2019.10.010},
   url = {https://www.sciencedirect.com/science/article/pii/S1000936119304170},
   year = {2021},
   type = {Journal Article}
}

@article{RN1639,
   author = {Ruud, C. O.},
   title = {A review of selected non-destructive methods for residual stress measurement},
   journal = {NDT International},
   volume = {15},
   number = {1},
   pages = {15-23},
   ISSN = {0308-9126},
   DOI = {https://doi.org/10.1016/0308-9126(82)90083-9},
   url = {https://www.sciencedirect.com/science/article/pii/0308912682900839},
   year = {1982},
   type = {Journal Article}
}

@article{RN145,
   author = {Hughes, D. S. and Kelly, J. L.},
   title = {Second-Order Elastic Deformation of Solids},
   journal = {Physical Review},
   volume = {92},
   number = {5},
   pages = {1145-1149},
   note = {PR},
   DOI = {10.1103/PhysRev.92.1145},
   url = {https://link.aps.org/doi/10.1103/PhysRev.92.1145},
   year = {1953},
   type = {Journal Article}
}

@article{RN1678,
   author = {Li, Bo and Cao, Yan-Ping and Feng, Xi-Qiao and Gao, Huajian},
   title = {Mechanics of morphological instabilities and surface wrinkling in soft materials: a review},
   journal = {Soft Matter},
   volume = {8},
   number = {21},
   pages = {5728-5745},
   ISSN = {1744-683X},
   DOI = {10.1039/C2SM00011C},
   url = {http://dx.doi.org/10.1039/C2SM00011C},
   year = {2012},
   type = {Journal Article}
}

@article{RN708,
   author = {De Belly, Henry and Yan, Shannon and Borja da Rocha, Hudson and Ichbiah, Sacha and Town, Jason P. and Zager, Patrick J. and Estrada, Dorothy C. and Meyer, Kirstin and Turlier, Hervé and Bustamante, Carlos and Weiner, Orion D.},
   title = {Cell protrusions and contractions generate long-range membrane tension propagation},
   journal = {Cell},
   volume = {186},
   number = {14},
   pages = {3049-3061.e15},
   ISSN = {0092-8674},
   DOI = {https://doi.org/10.1016/j.cell.2023.05.014},
   url = {https://www.sciencedirect.com/science/article/pii/S0092867423005330},
   year = {2023},
   type = {Journal Article}
}

@article{RN1027,
   author = {Maniou, Eirini and Todros, Silvia and Urciuolo, Anna and Moulding, Dale A. and Magnussen, Michael and Ampartzidis, Ioakeim and Brandolino, Luca and Bellet, Pietro and Giomo, Monica and Pavan, Piero G. and Galea, Gabriel L. and Elvassore, Nicola},
   title = {Quantifying mechanical forces during vertebrate morphogenesis},
   journal = {Nature Materials},
   ISSN = {1476-4660},
   DOI = {10.1038/s41563-024-01942-9},
   url = {https://doi.org/10.1038/s41563-024-01942-9},
   year = {2024},
   type = {Journal Article}
}

@article{RN854,
   author = {Firmin, Julie and Ecker, Nicolas and Rivet Danon, Diane and Özgüç, Özge and Barraud Lange, Virginie and Turlier, Hervé and Patrat, Catherine and Maître, Jean-Léon},
   title = {Mechanics of human embryo compaction},
   journal = {Nature},
   volume = {629},
   number = {8012},
   pages = {646-651},
   ISSN = {1476-4687},
   DOI = {10.1038/s41586-024-07351-x},
   url = {https://doi.org/10.1038/s41586-024-07351-x},
   year = {2024},
   type = {Journal Article}
}

@article{RN1514,
   author = {Alisafaei, Farid and Shakiba, Delaram and Hong, Yuan and Ramahdita, Ghiska and Huang, Yuxuan and Iannucci, Leanne E. and Davidson, Matthew D. and Jafari, Mohammad and Qian, Jin and Qu, Chengqing and Ju, David and Flory, Dashiell R. and Huang, Yin-Yuan and Gupta, Prashant and Jiang, Shumeng and Mujahid, Aliza and Singamaneni, Srikanth and Pryse, Kenneth M. and Chao, Pen-hsiu Grace and Burdick, Jason A. and Lake, Spencer P. and Elson, Elliot L. and Huebsch, Nathaniel and Shenoy, Vivek B. and Genin, Guy M.},
   title = {Tension anisotropy drives fibroblast phenotypic transition by self-reinforcing cell–extracellular matrix mechanical feedback},
   journal = {Nature Materials},
   volume = {24},
   number = {6},
   pages = {955-965},
   ISSN = {1476-4660},
   DOI = {10.1038/s41563-025-02162-5},
   url = {https://doi.org/10.1038/s41563-025-02162-5},
   year = {2025},
   type = {Journal Article}
}

@article{RN1546,
   author = {Jain, Akanksha and Gut, Gilles and Sanchis-Calleja, Fátima and Tschannen, Reto and He, Zhisong and Luginbühl, Nicolas and Zenk, Fides and Chrisnandy, Antonius and Streib, Simon and Harmel, Christoph and Okamoto, Ryoko and Santel, Malgorzata and Seimiya, Makiko and Holtackers, René and Rohland, Juliane K. and Jansen, Sophie Martina Johanna and Lutolf, Matthias P. and Camp, J. Gray and Treutlein, Barbara},
   title = {Morphodynamics of human early brain organoid development},
   journal = {Nature},
   volume = {644},
   pages = {1010-1019},
   ISSN = {1476-4687},
   DOI = {10.1038/s41586-025-09151-3},
   url = {https://doi.org/10.1038/s41586-025-09151-3},
   year = {2025},
   type = {Journal Article}
}

@article{RN1564,
   author = {Mulhall, Eric M. and Gharpure, Anant and Lee, Rachel M. and Dubin, Adrienne E. and Aaron, Jesse S. and Marshall, Kara L. and Spencer, Kathryn R. and Reiche, Michael A. and Henderson, Scott C. and Chew, Teng-Leong and Patapoutian, Ardem},
   title = {Direct observation of the conformational states of PIEZO1},
   journal = {Nature},
   volume = {620},
   number = {7976},
   pages = {1117-1125},
   ISSN = {1476-4687},
   DOI = {10.1038/s41586-023-06427-4},
   url = {https://doi.org/10.1038/s41586-023-06427-4},
   year = {2023},
   type = {Journal Article}
}

@article{RN1679,
   author = {Lv, Jian-Qing and Chen, Peng-Cheng and Chen, Yun-Ping and Liu, Hao-Yu and Wang, Shi-Da and Bai, Jianbo and Lv, Cheng-Lin and Li, Yue and Shao, Yue and Feng, Xi-Qiao and Li, Bo},
   title = {Active hole formation in epithelioid tissues},
   journal = {Nature Physics},
   volume = {20},
   number = {8},
   pages = {1313-1323},
   ISSN = {1745-2481},
   DOI = {10.1038/s41567-024-02504-1},
   url = {https://doi.org/10.1038/s41567-024-02504-1},
   year = {2024},
   type = {Journal Article}
}

@article{RN1033,
   author = {Shankar, Suraj and Mahadevan, L.},
   title = {Active hydraulics and odd elasticity of muscle fibres},
   journal = {Nature Physics},
   volume = {20},
   pages = {1501–1508},
   ISSN = {1745-2481},
   DOI = {10.1038/s41567-024-02540-x},
   url = {https://doi.org/10.1038/s41567-024-02540-x},
   year = {2024},
   type = {Journal Article}
}

@article{RN1605,
   author = {Ehret, Alexander E. and Böl, Markus and Itskov, Mikhail},
   title = {A continuum constitutive model for the active behaviour of skeletal muscle},
   journal = {Journal of the Mechanics and Physics of Solids},
   volume = {59},
   number = {3},
   pages = {625-636},
   ISSN = {0022-5096},
   DOI = {https://doi.org/10.1016/j.jmps.2010.12.008},
   url = {https://www.sciencedirect.com/science/article/pii/S0022509610002486},
   year = {2011},
   type = {Journal Article}
}

@article{RN1681,
   author = {Rogers, John A. and Someya, Takao and Huang, Yonggang},
   title = {Materials and Mechanics for Stretchable Electronics},
   journal = {Science},
   volume = {327},
   number = {5973},
   pages = {1603-1607},
   DOI = {doi:10.1126/science.1182383},
   url = {https://www.science.org/doi/abs/10.1126/science.1182383},
   year = {2010},
   type = {Journal Article}
}

@article{RN979,
   author = {Na, Hyeonuk and Kang, Yong-Woo and Park, Chang Seo and Jung, Sohyun and Kim, Ho-Young and Sun, Jeong-Yun},
   title = {Hydrogel-based strong and fast actuators by electroosmotic turgor pressure},
   journal = {Science},
   volume = {376},
   number = {6590},
   pages = {301-307},
   DOI = {doi:10.1126/science.abm7862},
   url = {https://www.science.org/doi/abs/10.1126/science.abm7862},
   year = {2022},
   type = {Journal Article}
}

@article{RN1682,
   author = {Theocharidis, Georgios and Yuk, Hyunwoo and Roh, Heejung and Wang, Liu and Mezghani, Ikram and Wu, Jingjing and Kafanas, Antonios and Contreras, Mauricio and Sumpio, Brandon and Li, Zhuqing and Wang, Enya and Chen, Lihong and Guo, Chuan Fei and Jayaswal, Navin and Katopodi, Xanthi-Leda and Kalavros, Nikolaos and Nabzdyk, Christoph S. and Vlachos, Ioannis S. and Veves, Aristidis and Zhao, Xuanhe},
   title = {A strain-programmed patch for the healing of diabetic wounds},
   journal = {Nature Biomedical Engineering},
   volume = {6},
   number = {10},
   pages = {1118-1133},
   ISSN = {2157-846X},
   DOI = {10.1038/s41551-022-00905-2},
   url = {https://doi.org/10.1038/s41551-022-00905-2},
   year = {2022},
   type = {Journal Article}
}

@article{RN669,
   author = {Gouveia, Bernardo and Kim, Yoonji and Shaevitz, Joshua W. and Petry, Sabine and Stone, Howard A. and Brangwynne, Clifford P.},
   title = {Capillary forces generated by biomolecular condensates},
   journal = {Nature},
   volume = {609},
   number = {7926},
   pages = {255-264},
   ISSN = {1476-4687},
   DOI = {10.1038/s41586-022-05138-6},
   url = {https://doi.org/10.1038/s41586-022-05138-6},
   year = {2022},
   type = {Journal Article}
}

@article{RN1627,
   author = {Verkest, Clement and Roettger, Lucas and Zeitzschel, Nadja and Hall, James and Sánchez-Carranza, Oscar and Huang, Angela Tzu-Lun and Lewin, Gary R. and Lechner, Stefan G.},
   title = {Cluster nanoarchitecture and structural diversity of PIEZO1 at rest and during activation in intact cells},
   journal = {Science Advances},
   volume = {11},
   number = {43},
   pages = {eady8052},
   note = {doi: 10.1126/sciadv.ady8052},
   DOI = {10.1126/sciadv.ady8052},
   url = {https://doi.org/10.1126/sciadv.ady8052},
   year = {2025},
   type = {Journal Article}
}

@article{RN1536,
   author = {Nia, Hadi T. and Munn, Lance L. and Jain, Rakesh K.},
   title = {Probing the physical hallmarks of cancer},
   journal = {Nature Methods},
   volume = {22},
   pages = {pages1800–1818},
   ISSN = {1548-7105},
   DOI = {10.1038/s41592-024-02564-4},
   url = {https://doi.org/10.1038/s41592-024-02564-4},
   year = {2025},
   type = {Journal Article}
}

@article{RN1498,
   author = {Jiang, Yuxuan and Li, Guo-Yang and Hu, Keshuai and Ma, Shiyu and Zheng, Yang and Jiang, Mingwei and Zhang, Zhaoyi and Wang, Xinyu and Cao, Yanping},
   title = {Simultaneous imaging of bidirectional guided waves probes arterial mechanical anisotropy, blood pressure, and stress synchronously},
   journal = {Science Advances},
   volume = {11},
   number = {32},
   pages = {eadv5660},
   DOI = {doi:10.1126/sciadv.adv5660},
   url = {https://www.science.org/doi/abs/10.1126/sciadv.adv5660},
   year = {2025},
   type = {Journal Article}
}

@article{RN1225,
   author = {Sleeboom, Jelle J. F. and van Tienderen, Gilles S. and Schenke-Layland, Katja and van der Laan, Luc J. W. and Khalil, Antoine A. and Verstegen, Monique M. A.},
   title = {The extracellular matrix as hallmark of cancer and metastasis: From biomechanics to therapeutic targets},
   journal = {Science Translational Medicine},
   volume = {16},
   number = {728},
   pages = {eadg3840},
   DOI = {doi:10.1126/scitranslmed.adg3840},
   url = {https://www.science.org/doi/abs/10.1126/scitranslmed.adg3840},
   year = {2024},
   type = {Journal Article}
}

@article{RN743,
   author = {Du, Yangkun and Lü, Chaofeng and Chen, Weiqiu and Destrade, Michel},
   title = {Modified multiplicative decomposition model for tissue growth: Beyond the initial stress-free state},
   journal = {Journal of the Mechanics and Physics of Solids},
   volume = {118},
   pages = {133-151},
   ISSN = {0022-5096},
   DOI = {https://doi.org/10.1016/j.jmps.2018.05.014},
   url = {https://www.sciencedirect.com/science/article/pii/S002250961830108X},
   year = {2018},
   type = {Journal Article}
}

@article{RN1612,
   author = {Cheng, Bo and Li, Moxiao and Lin, Min and Guo, Hui and Xu, Feng},
   title = {Mechanobiology across timescales},
   journal = {Nature Reviews Physics},
   ISSN = {2522-5820},
   DOI = {10.1038/s42254-025-00874-w},
   url = {https://doi.org/10.1038/s42254-025-00874-w},
   year = {2025},
   type = {Journal Article}
}

@article{RN1689,
   author = {Gómez-González, Manuel and Latorre, Ernest and Arroyo, Marino and Trepat, Xavier},
   title = {Measuring mechanical stress in living tissues},
   journal = {Nature Reviews Physics},
   volume = {2},
   number = {6},
   pages = {300-317},
   ISSN = {2522-5820},
   DOI = {10.1038/s42254-020-0184-6},
   url = {https://doi.org/10.1038/s42254-020-0184-6},
   year = {2020},
   type = {Journal Article}
}

@article{RN872,
   author = {Yin, Z. and Li, G. Y. and Zhang, Z. and Zheng, Y. and Cao, Y.},
   title = {SWENet: A Physics-Informed Deep Neural Network (PINN) for Shear Wave Elastography},
   journal = {IEEE Transactions on Medical Imaging},
   volume = {43},
   number = {4},
   pages = {1434-1448},
   ISSN = {1558-254X},
   DOI = {10.1109/TMI.2023.3338178},
   year = {2024},
   type = {Journal Article}
}

\subsection*{Acknowledgements}
This work is supported by the National Natural Science Foundation of China (Grant Nos. 12532016 and 12472176) and the Fundamental Research Funds for the Central Universities, Peking University.

\subsection*{Author contributions}
G.Y.L. designed research; Y.D. and G.Y.L. performed research; Y.D., M.D., and G.Y.L. analyzed data; all the authors wrote the paper.

\subsection*{Competing interests}
The authors declare no competing interests.


\clearpage
\section*{Supplementary Materials}
\renewcommand{\theequation}{S\arabic{equation}}
\renewcommand\thefigure{S\arabic{figure}}
\setcounter{equation}{0}
\setcounter{figure}{0}

\subsection*{Note1: Phase and group velocity}
Equation~\eqref{eq:5} in the main text, which governs elastic wave motion in incompressible soft solids under homogeneous prestress, can be written as
\begin{equation}\label{eq:s1.1}
{\mathcal A}_{p'i'q'j',p'}^0\,u_{j',q'} - {\hat{\bar{p}}_{,i'}} = \rho\, u_{i',tt},
\end{equation}
subject to the incompressibility constraint
\begin{equation}\label{eq:s1.2}
u_{i', i'} = 0 .
\end{equation}

We consider plane wave propagation in the principal plane \(x_{1'}\!-\!x_{2'}\).
In this case, the only nonzero components of the incremental elastic modulus tensor \(\mathcal{A}_{p'i'q'j'}^0\) entering the governing equations are
\(\mathcal{A}_{i'i'i'i'}^0\), \(\mathcal{A}_{i'i'j'j'}^0\), \(\mathcal{A}_{i'j'j'i'}^0\), and \(\mathcal{A}_{i'j'i'j'}^0\),
with \(i', j' \in \{1',2'\}\).
The equations of motion then reduce to
\begin{subequations}
\begin{align}
&{\mathcal A}_{1'1'1'1'}^0  {u_{1',1'1'}} 
+ {\mathcal A}_{2'1'2'1'}^0  {u_{1',2'2'}} 
+ {\mathcal A}_{2'1'1'2'}^0  {u_{2',2'1'}} 
+ {\mathcal A}_{1'1'2'2'}^0  {u_{2',1'2'}} 
- \hat{\bar{p}}_{,1'}
= \rho  {u_{1',tt}}, \label{eq:s1.3a} \\ 
&{\mathcal A}_{2'2'1'1'}^0  {u_{1',2'1'}} 
+ {\mathcal A}_{1'2'2'1'}^0  {u_{1',1'2'}} 
+ {\mathcal A}_{1'2'1'2'}^0  {u_{2',1'1'}} 
+ {\mathcal A}_{2'2'2'2'}^0  {u_{2',2'2'}} 
- \hat{\bar{p}}_{,2'} 
= \rho  {u_{2',tt}}. \label{eq:s1.3b}
\end{align}
\end{subequations}

To enforce incompressibility, we introduce a scalar stream function
\(\psi = \psi(x_{1'},x_{2'},t)\) such that
\[
u_{1'} = \psi_{,2'}, \qquad u_{2'} = -\psi_{,1'} .
\]
Substituting these expressions into Eq.~\eqref{eq:s1.3a} and~\eqref{eq:s1.3b}, and eliminating the Lagrange multiplier \(\hat{\bar{p}}\), yields
\begin{equation}\label{eq:s1.4}
\begin{split}
{\mathcal A}_{1'2'1'2'}^0  {\psi _{,1'1'1'1'}} 
&+ \left( {\mathcal A}_{1'1'1'1'}^0     
- 2{\mathcal A}_{1'2'2'1'}^0 
- 2{\mathcal A}_{1'1'2'2'}^0  
+ {\mathcal A}_{2'2'2'2'}^0 \right)  
{\psi _{,1'1'2'2'}} \\
&+ {\mathcal A}_{2'1'2'1'}^0  {\psi _{,2'2'2'2'}} 
= \rho \left( {\psi_{,2'2'tt}} +  {\psi_{,1'1'tt}} \right).
\end{split}
\end{equation}

For plane wave propagation, the stream function takes the form
\begin{equation}
\psi(x_{1'}, x_{2'})\mathrm{e}^{\mathrm{i}\omega t} 
= \psi_0 \exp \left[ \mathrm{i}k\left( x_{1'}\cos \vartheta 
+ x_{2'}\sin \vartheta - vt \right) \right],
\label{eq:s1.5}
\end{equation}
where \(\psi_0\) is the constant amplitude, \(\vartheta\) is the propagation angle measured from the \(x_{1'}\) axis, \(k\) is the wavenumber, and \(v\) is the phase velocity.

Substituting Eq.~\eqref{eq:s1.5} into~\eqref{eq:s1.4} yields the dispersion relation
\begin{equation}\label{eq:s1.6}
\rho v^2 
= \alpha \cos^4 \vartheta 
+ 2\beta \cos^2 \vartheta \sin^2 \vartheta 
+ \gamma \sin^4 \vartheta,
\end{equation}
where
\[
\alpha = {\mathcal A}_{1'2'1'2'}^0, \qquad
2\beta = {\mathcal A}_{1'1'1'1'}^0 
- 2{\mathcal A}_{1'2'2'1'}^0 
- 2{\mathcal A}_{1'1'2'2'}^0 
+ {\mathcal A}_{2'2'2'2'}^0, \qquad
\gamma = {\mathcal A}_{2'1'2'1'}^0 .
\]

Equation~\eqref{eq:s1.6} reveals the angular dependence of the phase velocity \(v\), which leads to wave dispersion.
The group velocity \(\boldsymbol v_g\), describing the propagation of a wave packet, is given by
\begin{equation}\label{eq:s1.7}
\boldsymbol v_g = \nabla_{\boldsymbol k}(kv),
\end{equation}
where \(\boldsymbol k = (k\cos \vartheta, k\sin \vartheta)\) is the wave vector.
Substituting the dispersion relation into Eq.~\eqref{eq:s1.7} yields
\begin{subequations}
\begin{align}
v_{g1} &= \frac{\alpha\cos \vartheta + 2\eta \sin^4 \vartheta \cos \vartheta}{\rho v}, \label{eq:s1.8a} \\
v_{g2} &= \frac{\gamma\sin \vartheta + 2\eta \sin \vartheta \cos^4 \vartheta}{\rho v}, \label{eq:s1.8b}
\end{align}    
\end{subequations}
where \(2\eta = 2\beta - \alpha - \gamma\).

\begin{figure*}[b]
\centering
\includegraphics{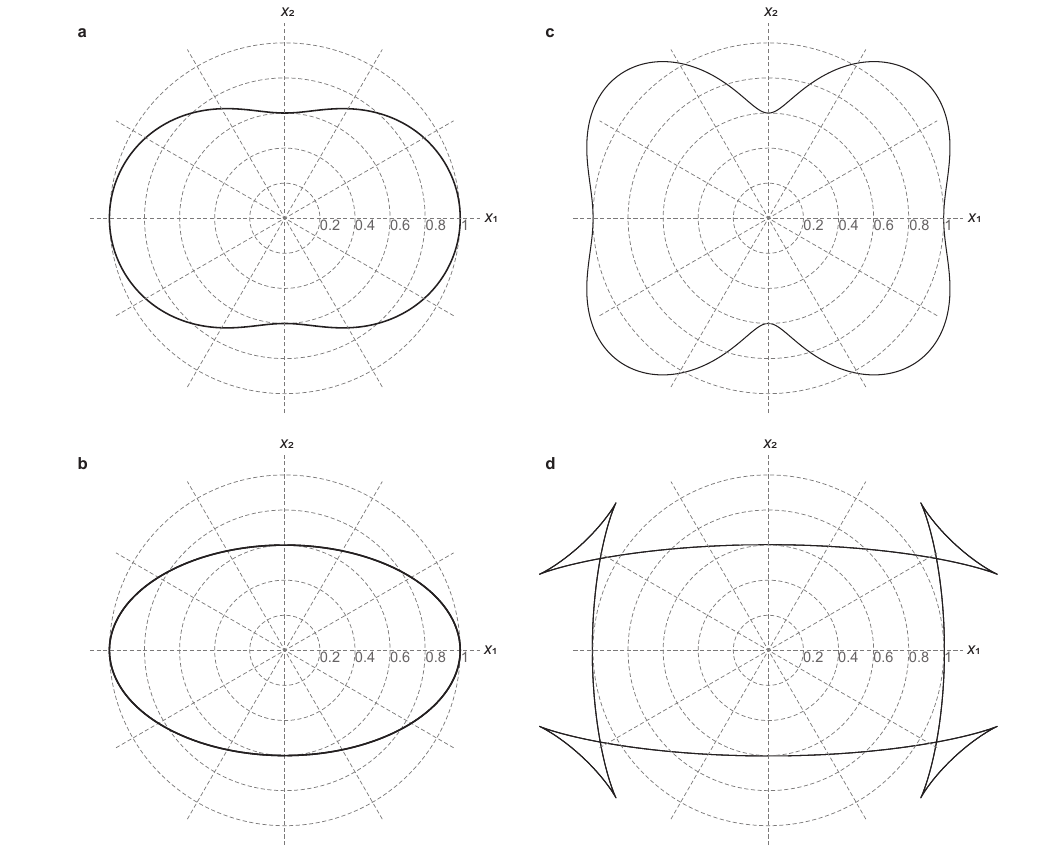}
\caption{Angle dependence of phase and group velocities in prestressed solids.
(a, b) Phase velocity and group velocity for \(\bar{\gamma}=0.36\) and \(\bar{\eta}=0\).
(c, d) Phase velocity and group velocity for \(\bar{\gamma}=0.36\) and \(\bar{\eta}=1.4\).}
\label{fig:S1}
\end{figure*}

To illustrate the angular dependence of the phase and group velocities, we introduce the dimensionless parameters
\(\bar{\gamma} = \gamma/\alpha\) and \(\bar{\eta} = \eta/\alpha\).
Figure~\ref{fig:S1} shows the phase and group velocities for the two cases
\((\bar{\gamma}, \bar{\eta}) = (0.36, 0)\) and \((0.36, 1.4)\).
For isotropic materials under moderate prestress, \(\bar{\eta} \approx 0\)~\cite{RN51}, and the group velocity exhibits an elliptical profile (Fig.~\ref{fig:S1}b).
In contrast, for anisotropic materials, \(\bar{\eta}\) can take large positive values, for example due to fiber reinforcement, leading to the emergence of cusps in the group velocity profile (Fig.~\ref{fig:S1}d). 

\subsection*{Note2: Coordinate transformations for the high-order derivates of the scalar steam function \(\psi\)}
The first-order partial derivatives of the stream function can be written as:
\begin{subequations}
\begin{align}
&{\psi _{,1'}} = {\psi _{,1}}\cos \theta  + {\psi _{,2}}\sin \theta, \\
&{\psi _{,2'}} =  - {\psi _{,1}}\sin \theta  + {\psi _{,2}}\cos \theta, \\
&{{\hat{\bar {p}}}_{,1'}} = {{\hat{\bar {p}}}_{,1'}}\cos \theta  + {{\hat{\bar {p}}}_{,2'}}\sin \theta, \\
&{{\hat{\bar {p}}}_{,2'}} =  - {{\hat{\bar {p}}}_{,1'}}\sin \theta  + {{\hat{\bar {p}}}_{,2'}}\cos \theta.
\end{align}    
\end{subequations}

The third-order partial derivatives of the stream function can be written as:
\begin{subequations}\label{eq:A9}
\begin{align}
{\psi _{,1'1'1'}} &= {\psi _{,111}}{\cos ^3}\theta  + 3{\psi _{,112}}{\cos ^2}\theta \sin \theta  \\
&+ 3{\psi _{,122}}\cos \theta {\sin ^2}\theta  + {\psi _{,222}}{\sin ^3}\theta, \\
{\psi _{,1'1'2'}} &=  - {\psi _{,111}}{\cos ^2}\theta \sin \theta  + {\psi _{,112}}\left( {{{\cos }^3}\theta  - 2\cos \theta {{\sin }^2}\theta } \right) \\
&+ {\psi _{,122}}\left( {2{{\cos }^2}\theta \sin \theta  - {{\sin }^3}\theta } \right) + {\psi _{,222}}\cos \theta {\sin ^2}\theta, \\
{\psi _{,1'2'2'}} &= {\psi _{,111}}\cos \theta {\sin ^2}\theta  - {\psi _{,112}}\left( {2{{\cos }^2}\theta \sin \theta  - {{\sin }^3}\theta } \right){\mkern 1mu}  \\
&+ {\psi _{,122}}\left( {{{\cos }^3}\theta  - 2\cos \theta {{\sin }^2}\theta } \right) + {\psi _{,222}}{\cos ^2}\theta \sin \theta, \\
{\psi _{,2'2'2'}} &=  - {\psi _{,111}}{\sin ^3}\theta  + 3{\psi _{,112}}\cos \theta {\sin ^2}\theta  \\
&- 3{\psi _{,122}}{\cos ^2}\theta \sin \theta  + {\psi _{,222}}{\cos ^3}\theta. 
\end{align}    
\end{subequations}

In the frequency domain, the physics-informed part of the neural network loss function can be obtained from the following equation:
\begin{subequations}\label{eq:A10}
\begin{align}
&{\beta _1} {\psi _{,1'1'2'}} + \alpha  {\psi _{,2'2'2'}} - {{\dot p}_{,1'}} + \rho {\omega ^2} {\psi _{,2'}} = 0,\\
&{\beta _2} {\psi _{,1'2'2'}} + \gamma  {\psi _{,1'1'1'}} +  {{\dot p}_{,2'}} + \rho {\omega ^2} {\psi _{,1'}} = 0.
\end{align}    
\end{subequations}

Note that \(\psi\) is a complex-valued function, and the governing equation is enforced such that both its real and imaginary parts vanish.

The data-driven part of the loss function can be obtained from the following equation:
\begin{subequations}\label{eq:A11}
\begin{align}
& {\psi _{,1'}} +  {u_{2'}} = 0,\quad  {\psi _{,2'}} -  {u_{1'}} = 0.
\end{align}    
\end{subequations}

\subsection*{Note3: Non-uniqueness of the coefficients for the wave equations}
\begin{figure*}[b]
\centering
\includegraphics[width=1\textwidth]{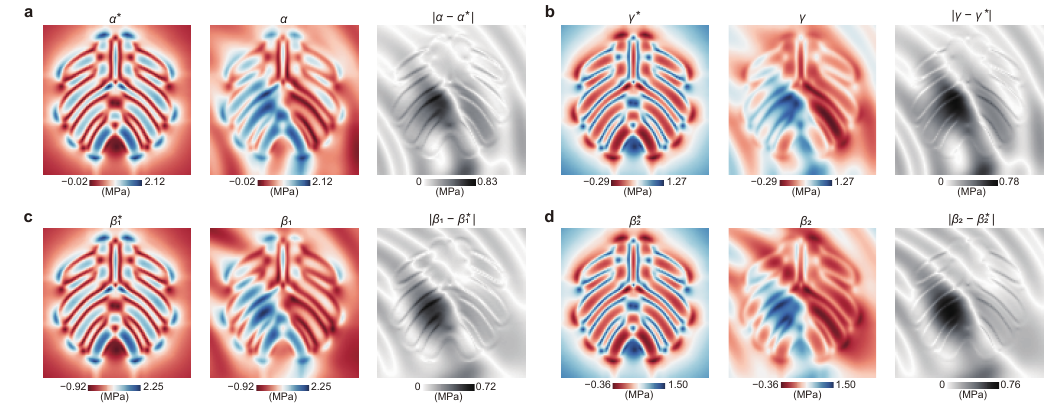}
\caption{
Comparison between inversion results and ground truth, along with their differences for \(\alpha\),  \(\gamma\),  \(\beta_1\) and  \(\beta_2\). Panels (a–d) correspond to \(\alpha\),  \(\gamma\),  \(\beta_1\) and  \(\beta_2\), respectively.}
\label{fig:s2}
\end{figure*}

We find that the inversion of $\alpha$, $\gamma$, $\beta_1$, and $\beta_2$ is not unique.
In particular, $\alpha$ and $\gamma$ are determined only up to an additive constant, as shown in Fig.~\ref{fig:s2}.

To illustrate this, we consider the case in which wave fields are measured in practice, from which the wave speeds in two perpendicular directions, $v_1$ and $v_2$, and the angle between the propagation direction and the principal direction, $\theta$, can be determined.
Since the relation
\begin{equation}\label{eq:A32}
\rho v_1^2 - \rho v_2^2 = (\alpha - \gamma)\cos 2\vartheta
\end{equation}
always holds, the quantity $\alpha - \gamma$ can be uniquely determined, and therefore the stress difference $\sigma_1 - \sigma_2 = \alpha - \gamma$ is uniquely identified.

However, $\alpha$ and $\gamma$ themselves are not uniquely determined, as they can be shifted by the same constant $c$, i.e., replaced by $\alpha + c$ and $\gamma + c$. In this case, Eq.~\eqref{eq:A32} remains unchanged.
Consequently, the inversion cannot distinguish between the true solution and a shifted solution in which both $\alpha$ and $\gamma$ differ from their true values by the same constant.
This demonstrates that $\alpha$ and $\gamma$ are identifiable only up to a common additive constant, as evidenced by the results in Fig.~\ref{fig:s2}.

Similarly, $\beta_1$ and $\beta_2$ are also not uniquely identifiable and can differ by a shift that depends on the constant $c$, which is related to the displacement field:
\begin{subequations}\label{eq:A34}
\begin{align}
&b_1 = -c \dfrac{\sin^2 \vartheta}{\cos^2 \vartheta}, \\
&b_2 = -c \dfrac{\cos^2 \vartheta}{\sin^2 \vartheta}.
\end{align}
\end{subequations}

\subsection*{Note4: Stress identification for compressible materials}
For incompressible materials, the wave equation should be written as:
\begin{subequations}\label{eq:s21}
\begin{align}
{\mathcal A_{1'1'1'1'}^0}{u_{1',1'1'}} + ({\mathcal A_{2'1'1'2'}^0} + {\mathcal A_{1'1'2'2'}^0}){u_{2',1'2'}} + {\mathcal A_{2'1'2'1'}^0}{u_{1',2'2'}} = \rho {\omega ^2}{u_{1'}},\\
{\mathcal A_{1'2'1'2'}^0}{u_{2',1'1''}} + ({\mathcal A_{2'1'1'2'}^0} + {\mathcal A_{1'1'2'2'}^0}){u_{1',1'2'}} + {\mathcal A_{2'2'2'2'^0}}{u_{2',2'2'}} = \rho {\omega ^2}{u_{2'}}.
\end{align}
\end{subequations}

Let \(\alpha  = {\mathcal A_{1'2'1'2'}^0}, \quad \gamma  = {\mathcal A_{2'1'2'1'}^0}, \quad {\beta _1} = {\mathcal A_{1'1'1'1'}^0}, \quad {\beta _2} = {\mathcal A_{2'2'2'2'}^0}, \quad {\beta _3} = {\mathcal A_{2'1'1'2'}^0} + {\mathcal A_{1'1'2'2'}^0}\).
Assume a compressible constitutive model of the following form:
\begin{equation}
W = W\left( {{I_1},{I_2}} \right) + \frac{1}{{{D_1}}}{\left( {\sqrt {{I_3}}  - 1} \right)^2},
\label{eq:s22}
\end{equation}
in which the volumetric deformation term is fixed as \(\frac{1}{{{D_1}}}{\left( {\sqrt {{I_3}}  - 1} \right)^2}\), where \({{D_1}}\) is related to the bulk modulus.

Under the plane strain assumption, consider a biaxial stretch \({\bf{F}} = {\mathrm{diag}}\left( {{\lambda _1},{\lambda _2},1} \right)\). The associated invariants can be expressed as
\begin{subequations}\label{eq:s23}
\begin{align}
&{I_1} = \lambda _1^2 + \lambda _2^2 + 1,\\
&{I_2} = \lambda _1^2\lambda _2^2 + \lambda _2^2 + \lambda _1^2,\\
&{I_3} = \lambda _1^2\lambda _2^2.
\end{align}
\end{subequations}

Substituting these into the constitutive model yields:
\begin{equation}
{\bf{\sigma }} = \frac{2}{J}\left[ {{W_1}{\bf{B}} + {W_2}\left( {{I_1}{\bf{B}} - {{\bf{B}}^2}} \right)} \right] + \frac{2}{{{D_1}}}\left( {J - 1} \right){\bf{I}},
\label{eq:s24}
\end{equation}
where \({W_1} = \frac{{\partial W}}{{\partial {I_1}}}\), \({W_2} = \frac{{\partial W}}{{\partial {I_2}}}\), \({W_{11}} = \frac{{{\partial ^2}W}}{{\partial I_1^2}}\), \({W_{12}} = \frac{{{\partial ^2}W}}{{\partial {I_1}\partial {I_2}}}\), \({W_{22}} = \frac{{{\partial ^2}W}}{{\partial I_2^2}}\), \({\bf{B}} = {\bf{F}}{{\bf{F}}^T}.\)

The tensor \({\mathcal A}_{0piqj}\) can be expressed as:
\begin{subequations}\label{eq:S25}
\begin{align}
&{\mathcal A_{i'i'i'i'}^0} = \frac{{\lambda _i^2}}{J}\frac{{{\partial ^2}W}}{{\partial \lambda _i^2}},\quad{\mathcal A_{i'i'j'j'}^0} = \frac{{{\partial ^2}W}}{{\partial {\lambda _i}\partial {\lambda _j}}},\\
&{\mathcal A_{i'j'j'i'}^0} = \frac{1}{{\lambda _i^2 - \lambda _j^2}}\left( {{\lambda _j}\frac{{\partial W}}{{\partial {\lambda _i}}} - {\lambda _i}\frac{{\partial W}}{{\partial {\lambda _j}}}} \right),\\
&{\mathcal A_{i'j'i'j'}^0} = \frac{{\lambda _i^2}}{{J\left( {\lambda _i^2 - \lambda _j^2} \right)}}\left( {{\lambda _i}\frac{{\partial W}}{{\partial {\lambda _i}}} - {\lambda _j}\frac{{\partial W}}{{\partial {\lambda _j}}}} \right).
\end{align}
\end{subequations}

According to the chain rule, the partial derivatives of the strain energy density with respect to the stretches satisfy the following relations:
\begin{subequations}\label{eq:s26}
\begin{align}
&\frac{{\partial W}}{{\partial {\lambda _1}}} = 2{\lambda _1}\left[ {{W_1} + \left( {\lambda _2^2 + 1} \right){W_2}} \right] + \frac{{2{\lambda _2}}}{{{D_1}}}\left( {J - 1} \right),\\
&\frac{{\partial W}}{{\partial {\lambda _2}}} = 2{\lambda _2}\left[ {{W_1} + \left( {\lambda _1^2 + 1} \right){W_2}} \right] + \frac{{2{\lambda _1}}}{{{D_1}}}\left( {J - 1} \right),\\
&\frac{{{\partial ^2}W}}{{\partial \lambda _1^2}} = 2\left[ {{W_1} + \left( {\lambda _2^2 + 1} \right){W_2} + \frac{{\lambda _2^2}}{{{D_1}}}\left( {1 - \frac{1}{J}} \right)} \right] + 4\lambda _1^2\left[ {{W_{11}} + 2\left( {\lambda _2^2 + 1} \right){W_{12}} + {{\left( {\lambda _2^2 + 1} \right)}^2}{W_{22}}} \right] + \frac{{2\lambda _2^2}}{{{D_1}J}},\\
&\frac{{{\partial ^2}W}}{{\partial \lambda _2^2}} = 2\left[ {{W_1} + \left( {\lambda _1^2 + 1} \right){W_2} + \frac{{\lambda _1^2}}{{{D_1}}}\left( {1 - \frac{1}{J}} \right)} \right] + 4\lambda _2^2\left[ {{W_{11}} + 2\left( {\lambda _1^2 + 1} \right){W_{12}} + {{\left( {\lambda _1^2 + 1} \right)}^2}{W_{22}}} \right] + \frac{{2\lambda _1^2}}{{{D_1}J}},\\
&\frac{{{\partial ^2}W}}{{\partial {\lambda _1}\partial {\lambda _2}}} = 4{\lambda _1}{\lambda _2}{W_2} + \frac{4}{{{D_1}}}\left( {J - 1} \right) + 4{\lambda _1}{\lambda _2}\left[ {{W_{11}} + \left( {\lambda _1^2 + \lambda _2^2 + 2} \right){W_{12}} + \left( {\lambda _1^2 + 1} \right)\left( {\lambda _2^2 + 1} \right){W_{22}}} \right] + \frac{2}{{{D_1}}}.
\end{align}    
\end{subequations}

Substituting these expressions, we obtain:
\begin{subequations}\label{eq:s27}
\begin{align}
&\alpha  = 2\left( {{W_1} + {W_2}} \right)\frac{{{\lambda _1}}}{{{\lambda _2}}},\quad\gamma  = 2\left( {{W_1} + {W_2}} \right)\frac{{{\lambda _2}}}{{{\lambda _1}}},\\
&{\beta _1} = 2\left[ {{W_1} + \left( {\lambda _2^2 + 1} \right){W_2}} \right]\frac{{{\lambda _1}}}{{{\lambda _2}}} + 4\lambda _1^2\left[ {{W_{11}} + 2\left( {\lambda _2^2 + 1} \right){W_{12}} + {{\left( {\lambda _2^2 + 1} \right)}^2}{W_{22}}} \right]\frac{{{\lambda _1}}}{{{\lambda _2}}} + \frac{{2J}}{{{D_1}}},\\
&{\beta _2} = 2\left[ {{W_1} + \left( {\lambda _1^2 + 1} \right){W_2}} \right]\frac{{{\lambda _2}}}{{{\lambda _1}}} + 4\lambda _2^2\left[ {{W_{11}} + 2\left( {\lambda _1^2 + 1} \right){W_{12}} + {{\left( {\lambda _1^2 + 1} \right)}^2}{W_{22}}} \right]\frac{{{\lambda _2}}}{{{\lambda _1}}} + \frac{{2J}}{{{D_1}}},\\
&{\beta _3} = 2J{W_2} + 4J\left[ {{W_{11}} + \left( {\lambda _1^2 + \lambda _2^2 + 2} \right){W_{12}} + \left( {\lambda _1^2 + 1} \right)\left( {\lambda _2^2 + 1} \right){W_{22}}} \right] + \frac{{2J}}{{{D_1}}}.
\end{align}
\end{subequations}

In particular, for the neo-Hookean model:
\begin{subequations}\label{eq:s28}
\begin{align}
&\alpha  = 2{c_1}\frac{{{\lambda _1}}}{{{\lambda _2}}},\quad\gamma  = 2{c_1}\frac{{{\lambda _2}}}{{{\lambda _1}}},\quad{\beta _1} = 2{c_1}\frac{{{\lambda _1}}}{{{\lambda _2}}} + \frac{{2J}}{{{D_1}}},\\
&{\beta _2} = 2{c_1}\frac{{{\lambda _2}}}{{{\lambda _1}}} + \frac{{2J}}{{{D_1}}},\quad{\beta _3} = \frac{{2J}}{{{D_1}}}.
\end{align}
\end{subequations}

\clearpage

\clearpage

\clearpage
\begin{figure*}[b]
\centering
\includegraphics[width=1\textwidth]{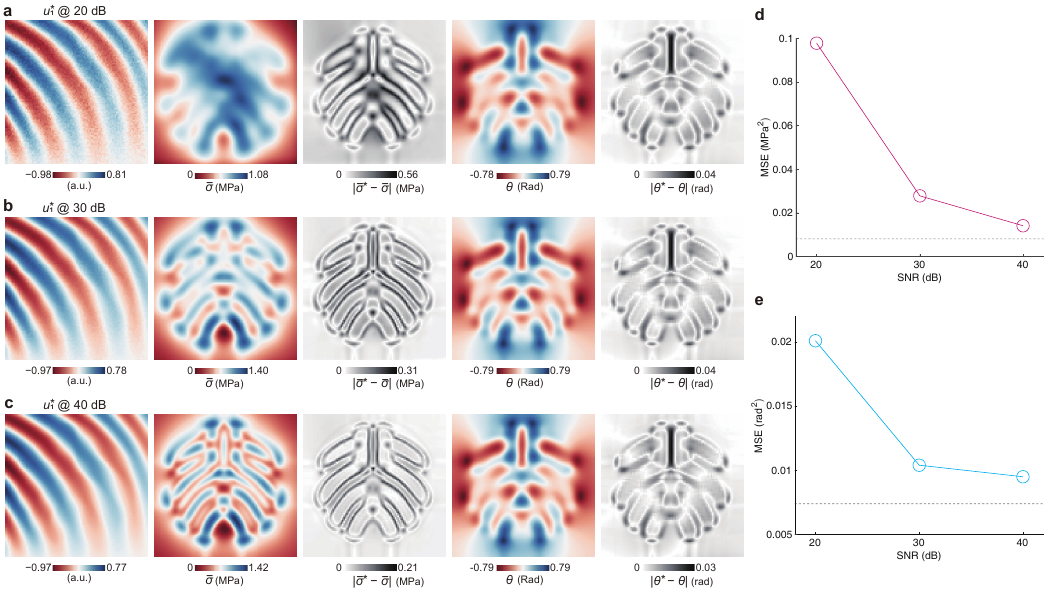}
\caption{
(a)-(c) Wave fields and inversion results for SNRs of 20, 30, and 40 dB.  
(d) Mean squared error curves of the inversion results under different SNRs.
}
\label{fig:s3}
\end{figure*}

\clearpage
\begin{figure*}[b]
\centering
\includegraphics[width=1\textwidth]{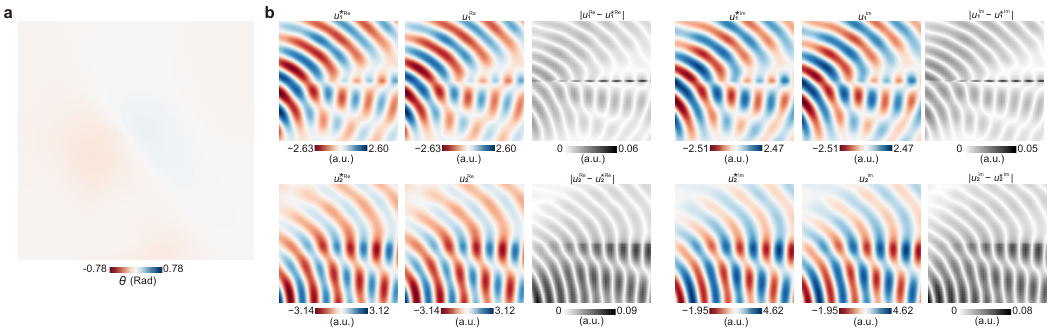}
\caption{
(a) Inverted principal stress directions.  
(b) Shear-wave field in a layered normal stress field.
}
\label{fig:s4}
\end{figure*}

\clearpage
\begin{figure*}[b]
\centering
\includegraphics[width=1\textwidth]{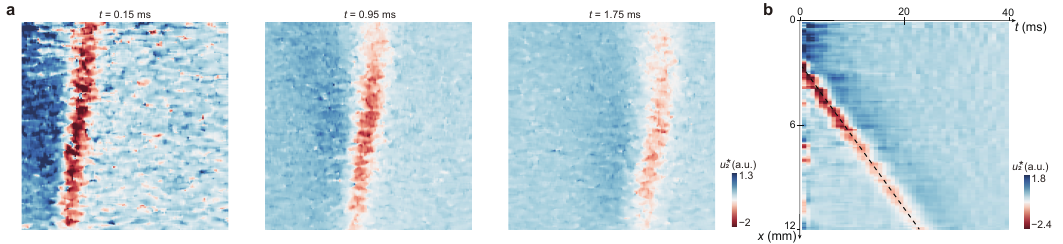}
\caption{
(a) Snapshots showing the  shear wave propagation in the ROI. The maps depict the vertical particle velocity fields. 
(b) Spatiotemporal maps of the horizontally propagated shear waves, where we get a shear wave speed of 4.3 m/s.
}
\label{fig:s5}
\end{figure*}

\end{document}